\documentclass[ALICE,manyauthors]{cernphprep}

\usepackage{rotating}
\usepackage[normalem]{ulem}

\usepackage{amssymb,amsmath,amsfonts}
\usepackage{graphicx}
\usepackage{amssymb}
\usepackage{epstopdf}
\usepackage{cite}
\usepackage{multirow}
\usepackage{color}
\usepackage{amsmath,bm}
\usepackage{amsbsy}
\usepackage{ulem}
\usepackage{siunitx}

\usepackage{hyperref}

\usepackage{datetime}
\usdate
\usepackage{lineno}
\usepackage[toc,page]{appendix}
\usepackage[section] {placeins}

\usepackage[T1]{fontenc}

\newcommand{\CommentBlock}[1]{}

\newcommand{\sqrtsNN}{\ensuremath{\sqrt{s_{\mathrm {NN}}}}}
\newcommand{\sqrts}{\ensuremath{\sqrt{s}}}

\newcommand{\gev}{\ensuremath{\mathrm{GeV/}c}}

\newcommand{\kT}{\ensuremath{k_\mathrm{T}}}

\newcommand{\pT}{\ensuremath{p_\mathrm{T}}}

\newcommand{\pTjetch}{\ensuremath{p_\mathrm{T,jet}^\mathrm{ch}}}
\newcommand{\pTtrig}{\ensuremath{p_{\mathrm{T,trig}}}}

\newcommand{\TTSig}{\ensuremath{\mathrm{TT}_{\mathrm{Sig}}}}
\newcommand{\TTRef}{\ensuremath{\mathrm{TT}_{\mathrm{Ref}}}}

\newcommand{\DeltaR}{\ensuremath{\Delta{R}}}

\begin{document}
%\linenumbers

\begin{titlepage}
\PHnumber{082}                 % required, obtained from PH
\PHdate{10 May}              % required
\PHyear{2021}              % required

\title{First measurements of $N$-subjettiness in central Pb--Pb collisions at ${\sqrt{\it{s}_{\rm {NN}}}} = 2.76$ TeV}
\ShortTitle{First measurements of $N$-subjettiness in central Pb--Pb collisions} 

\Collaboration{ALICE Collaboration%
         \thanks{See Section~\ref*{app:collab} for the list of collaboration
                      members}}
\ShortAuthor{ALICE Collaboration}      % appears on left page headers, do not change

%=================================================================
\begin{abstract}
  
The ALICE Collaboration reports the first fully-corrected measurements of the $N$-subjettiness observable for track-based jets in heavy-ion collisions. This study is performed using data recorded in pp and Pb--Pb collisions at centre-of-mass energies of $\sqrts~= 7$ TeV and $\sqrt{s_{\rm NN}} = 2.76$\,TeV, respectively. In particular the ratio of 2-subjettiness to 1-subjettiness, $\tau_{2}/\tau_{1}$, which is sensitive to the rate of two-pronged jet substructure, is presented. Energy loss of jets traversing the strongly interacting medium in heavy-ion collisions is expected to change the rate of two-pronged substructure relative to vacuum.  The results are presented for jets with a resolution parameter of $R = 0.4$ and charged jet transverse momentum of $40 \leq p_{\rm T,\rm jet} \leq 60$ GeV/$c$, which constitute a larger jet resolution and lower jet transverse momentum interval than previous measurements in heavy-ion collisions. This has been achieved by utilising a semi-inclusive hadron-jet coincidence technique to suppress the larger jet combinatorial background in this kinematic region. No significant modification of the $\tau_{2}/\tau_{1}$ observable for track-based jets in Pb--Pb collisions is observed relative to vacuum PYTHIA6 and PYTHIA8 references at the same collision energy. The measurements of $\tau_{2}/\tau_{1}$, together with the splitting aperture angle $\Delta R$, are also performed in pp collisions at $\sqrt{s}=7$ TeV for inclusive jets. These results are compared with PYTHIA calculations at $\sqrt{s}=7$ TeV, in order to validate the model as a vacuum reference for the Pb--Pb centre-of-mass energy. The PYTHIA references for $\tau_{2}/\tau_{1}$ are shifted to larger values compared to the measurement in pp collisions. This hints at a reduction in the rate of two-pronged jets in Pb--Pb collisions compared to pp collisions.

\end{abstract}
\end{titlepage}
 
\setcounter{page}{2} %please do not remove this line

%\tableofcontents
%\setcounter{tocdepth}{3}

%=================================================================
\section{Introduction}
\label{sect:intro}

The goal of relativistic heavy-ion physics is to study the behaviour of Quantum Chromo-Dynamics (QCD) matter in the high energy density and temperature regimes, where a medium of deconfined quarks and gluons, known as the quark--gluon plasma (QGP), is formed\cite{Muller:2006ee,Roland:2014jsa}. During the initial stages of the collisions, short distance scattering interactions between the constituents of the incoming nucleons produce high-momentum transfer partons (quarks and gluons), which fragment into collimated showers of particles known as jets. Jets provide an experimental tool with which to reconstruct the parton shower and access the kinematics of the initial scattered parton. The production of jets and their substructure in pp collisions is well described by perturbative QCD~\cite{Dasgupta:2016bnd,Marzani:2019hun,Sirunyan:2018asm,Larkoski:2017jix}, which makes them well-calibrated probes to investigate this medium in heavy-ion collisions.  The phenomenon of jet quenching, which refers to the modification of jet production rates and jet substructure due to interactions in coloured matter~\cite{Majumder:2010qh}, is one of the most important signatures of QGP formation and has been extensively studied in nuclear collisions at both the Relativistic Heavy Ion Collider and the Large Hadron Collider, via measurements of inclusive jet production as well as correlations of jets with high-energy triggers (hadrons, photons, W and Z bosons and jets)~\cite{Adam:2015doa,Adamczyk:2017yhe,Sirunyan:2017jic,Aaboud:2018anc}. The observed modifications are a consequence of radiative and elastic interactions of the partons with the coloured medium and provide an opportunity for measuring the transport properties of the QGP, notably the transport parameter $\hat{q}$~\cite{Burke:2013yra}. However, the dynamical mechanisms underlying such interactions are still not well understood. Measurements of jet shapes~\cite{Acharya:2018uvf,Chatrchyan:2013kwa,Sirunyan:2018qec,Aaboud:2017bzv,Acharya:2017goa,Sirunyan:2018gct} and more recently, jet substructure ~\cite{Sirunyan:2017bsd,Kauder:2017mhg,Acharya:2019djg} attempt to shed light on these mechanisms. It has also been proposed that jet-medium interactions might be influenced by the underlying structure of the jet and the degree to which the medium can resolve this substructure~\cite{CasalderreySolana:2012ef}. Jet substructure observables can be used to quantify these modifications and study the responsible energy loss mechanisms, whilst also being sensitive to the fundamental properties of the medium~\cite{D'Eramo:2012jh}.

In this paper, modifications to the two-pronged substructure of jets in heavy-ion collisions, compared to those measured in pp collisions, are reported, in order to investigate the impact of the underlying jet substructure on quenching effects. Two-pronged jets are composed of exactly two distinct hard or semi-hard substructures and are tagged through the measurements of the ratio of 2-subjettiness to 1-subjettiness, $\tau_{2}/\tau_{1}$, which are calculated relative to a variety of differently defined axes. These are the axes of subjets obtained by reclustering the jet constituents with jet finding algorithms. Each algorithm can operationally reorder the jet shower, in momentum and angular separation, resulting in the selection of different axes. The reclustering algorithms are also combined with grooming techniques~\cite{Butterworth:2008iy,Larkoski:2014wba} to remove wide-angle soft radiation from the jet and select specific regions of the splitting phase space where medium-induced signals are enhanced (or suppressed) or where jet quenching calculations are under better control\cite{Andrews:2018jcm}.  In Pb--Pb collisions a semi-inclusive data driven approach is used to suppress the combinatorial background and extend the measurement to larger jet resolution and lower $p_{\rm T,\rm jet}$. In pp collisions, comparisons of the measurement with PYTHIA\cite{Sjostrand:2006za} calculations at the same centre-of-mass energy are used to validate the model as a vacuum reference for the Pb--Pb centre-of-mass energy. In addition to $\tau_{2}/\tau_{1}$, the splitting aperture angle, $\Delta R$, is also reported in pp collisions, providing complementary information on the underlying two-pronged substructure.

The paper is organised as follows: Section~\ref{sect:definition} describes the observables and the different axis choices (reclustering algorithms). Section~\ref{sect:Dataset} details the pp and Pb--Pb data sets and event selection criteria used for the analysis. Section~\ref{sect:JetReco} describes the jet finding procedure, including the underlying event and combinatorial background subtraction techniques used in Pb--Pb collisions. The response of the shapes to detector effects and background fluctuations are also highlighted. Section~\ref{sect:Unfolding} details the simultaneous correction of the jet substructure observable and $p_{\rm T,\rm jet}^{\rm ch}$ distributions, via a two-dimensional Bayesian unfolding procedure, for these detector effects and background fluctuations. Section~\ref{sect:sysuncert} lists the contributions to the systematic uncertainties. Finally the fully corrected results are presented in Section~\ref{sect:results}, alongside a discussion with comparisons to theoretical models. 
% =============================================================
\section{$N$-subjettiness, aperture angle and axis definitions}
 \label{sect:definition}
 This analysis measures the proportion of two-pronged jets in Pb--Pb compared to pp collisions. Track-based jet finding is performed using the anti-$k_{\rm{T}}$~\cite{FastJetAntikt} algorithm with a jet resolution parameter of $R=0.4$ (full details of the jet reconstruction are given in section~\ref{sect:JetReco}). In order to tag jets as being single-pronged or two-pronged, the $N$-subjettiness~\cite{Thaler:2010tr} observable is chosen. $N$-subjettiness is a jet substructure observable, denoted by $\tau_{N}$, which quantifies the degree to which a jet has a $N$(or fewer)-pronged substructure. It is measured relative to $N$ axes, which are the axes of the subjets returned by unwinding the reclustering history of a given choice of reclustering algorithm by $N-1$ steps, and is defined as,
 \begin{equation}
  \tau_{N}=\frac{1}{p_{\rm{T},\rm{jet}} \times R} \sum_{k} p_{\mathrm{T},\it{k}} \rm{ } \: \textrm{minimum} (\Delta \it{R}_{\mathrm{1},k},\rm{} \Delta \it{R}_{\mathrm{2},k},...., \rm{} \Delta \it{R}_{\mathrm{N},k}),
 \end{equation}
 
 where $k$ runs over the list of jet constituents. The transverse momentum, relative to the beam, of constituent $k$ is denoted as $p_{\mathrm{T},k}$ and $\Delta R_{\mathrm{S},k}$ is the distance in the pseudorapidity-azimuthal ($\eta$-$\varphi$) plane between the constituent $k$ and the axis of subjet $S$. The observable is normalised by the product of the jet resolution parameter, $R$, and the jet transverse momentum, $p_{\rm T,\rm jet}$.

If the bulk of the $p_{\rm T,\rm jet}$ is correlated to at least one of the subjet axes, the jet is composed of $N$ or fewer well defined subjets and $\tau_{N}$ tends to zero. If a sizeable fraction of the $p_{\rm T,\rm jet}$ is not aligned with any of the subjet axes, the jet is composed of at least $N+1$ subjets and $\tau_{N}$ tends to unity. The ratio of $\tau_{N}/\tau_{N-1}$ is sensitive to exactly $N$-prongs in a jet, as an $N$-pronged jet is expected to have low $\tau_{N}$ and high $\tau_{N-1}$ values. In this way, the ratio of the two quantities is more discriminative of $N$-prongness in jets than either quantity on its own.
 
 The $N$-subjettiness observable was originally designed to identify boosted hadronically-decaying objects such as W bosons and top quarks\cite{CMS-PAS-JME-13-007,Thaler:2010tr}  Reconstructed jets containing a W boson exhibit a distinct two-pronged energy flow due to the two hard subjets produced by the decay of the W boson to two quarks. The ratio of $\tau_{2}/\tau_{1}$ can be used to discriminate these jets from quark and gluon-initiated jets, which are primarily single-cored. In this paper the measured ratio of $\tau_{2}/\tau_{1}$, on a jet-by-jet basis, is used to identify the two-pronged subsample of QCD jets in both pp and Pb--Pb collisions. Jets with a clear two-pronged substructure relative to the subjet axes will have low $\tau_{2}$ and high $\tau_{1}$ values, resulting in a small $\tau_{2}/\tau_{1}$ ratio. Various jet quenching mechanisms, such as medium-induced semi-hard radiation emitted at large angles, are expected to change the rate of two-pronged QCD jets in heavy-ion collisions relative to the vacuum~\cite{Mehtar-Tani:2016aco}. Hard medium-induced radiation could be a signature of the jet interacting with the partonic structure of the QGP, since large momentum transfers are suppressed for strongly-coupled degrees of freedom\cite{D'Eramo:2012jh}. This type of radiation could create an additional prong in the jet, transforming the predominantly single-pronged QCD jets, into two-pronged objects. It has also been postulated that colour coherence effects~\cite{CasalderreySolana:2012ef}, arising from the finite resolving power of the medium with respect to jet substructure, could result in a larger degree of quenching for two-pronged jets compared to single-cored jets. This would result in a decrease in the population of two-pronged jets in Pb--Pb compared to pp collisions.

In addition to the  $\tau_{2}/\tau_{1}$ observable, the aperture angle between the two selected subjet axes in the $N=2$ case, $\Delta R$, is reported in pp collisions. Since the degree to which the jet substructure is two-pronged depends on the angular separation of the prongs, in addition to the way the $p_{\rm T,\rm jet}$ is distributed among them, this observable provides complementary information to $\tau_{2}/\tau_{1}$. The measurements of the $\Delta R$ observable in pp collisions also provide an important baseline for measurements of this observable in heavy-ion collisions, where they can be used to directly probe the angular resolving power of the medium with respect to coloured structures.

\subsection{ Subjet axes}
The calculation of $N$-subjettiness requires $N$ subjet axes, which are themselves obtained by unwinding the reclustering history $N-1$ steps. The subjet axes are therefore dependent on the choice of the reclustering algorithm used, with different algorithms returning axes that are sensitive to different regions of the splitting phase space. The addition of grooming techniques can further isolate this probed phase space, allowing for the selection of regions where calculations are under better theoretical control. The reclustering algorithms employed in this analysis, which belong to the sequential recombination class of algorithms, are detailed as follows:

\subsubsection{$k_{\rm{T}}$ clustering}
This metric clusters particles based on their $p_{\rm T}$ and angular separation from one another~\cite{Ellis:1993tq}. The clustering begins by combining soft particles, with hard structures only being brought together in the final steps of the clustering history. Therefore unwinding the last clustering step gives operational access to two hard subjets. 

\subsubsection{Cambridge-Aachen clustering}
The C/A algorithm combines particles solely based on their angular separation from one another, thus maintaining an angular ordered tree in vacuum~\cite{Dokshitzer:1997in}. Particles closest to each other in the $\eta$-$\varphi$ plane are brought together first, with the last step of the clustering combining the furthest separated structures in the jet. Unwinding the reclustering history one step gives access to the pair of subjets separated by the largest angle.

\subsubsection{Soft Drop grooming (with C/A)}
The Soft Drop groomer~\cite{Larkoski:2014wba} is applied to the C/A reclustered jet. In this case, instead of selecting the axes returned by unwinding the reclustering history by one step, the splitting is first tested against the Soft Drop condition. This is given by,

 \begin{equation}
   \frac{\textrm{minimum}(p_{\rm{T},1},p_{\rm{T},2})}{p_{\rm{T},1}+p_{\rm{T},2}} > z_{cut} \left(\frac{\Delta \it{R}}{R}\right)^{\beta},
   \label{req:SoftDrop}
 \end{equation}

where the indices $1$ and $2$ label the subjet axes and $R$ represents the jet resolution parameter. The $z_{cut}$ and $\beta$ parameters control the grooming behaviour. The former places a lower limit on the momentum fraction carried by the subleading subjet, which is the subjet carrying the smaller momentum fraction. The latter quantifies the interplay between the angular and momentum components of the splitting, with higher $\beta$ values preferentially removing soft large-angle radiation. If the Soft Drop condition is satisfied, the splitting is accepted and the axes are returned. However, if the splitting does not satisfy the Soft Drop condition, the subleading subjet is groomed away and the procedure is repeated by unclustering the leading subjet and testing against the Soft Drop condition. This process iteratively continues until either a splitting is found which passes the Soft Drop condition (the axes are returned), or the reclustering history is exhausted and there is only one track left in which case the jet is discarded.

The purpose of the grooming procedure is to strip away soft radiation from the jet and uncover a hard substructure. The benefit of using the C/A algorithm for the groomer is that the splitting tree is angular ordered, which is the expected configuration of the parton shower in the vacuum. Therefore, the first structure that is accepted by the groomer is expected to be correlated to the earliest hard splitting in the jet. In this work the Soft Drop condition is defined with values of $z_{cut}=0.1$ and $\beta=0$, which is equivalent to the Mass Drop tagger ~\cite{Dasgupta:2013ihk}. This removes the angular dependency from Eq.~\ref{req:SoftDrop} and reduces the Soft Drop condition to a test of the hardness of the splitting. In this configuration, the groomer removes soft large-angle radiation from the jet. The widest hard or semi-hard splitting that is found in the declustering process, is returned (in vacuum).

\subsubsection{Reclustering metric comparisons}

\begin{figure}[]
\centering
\begin{minipage}[b]{0.49\textwidth}
\centering
\includegraphics[width=\textwidth]{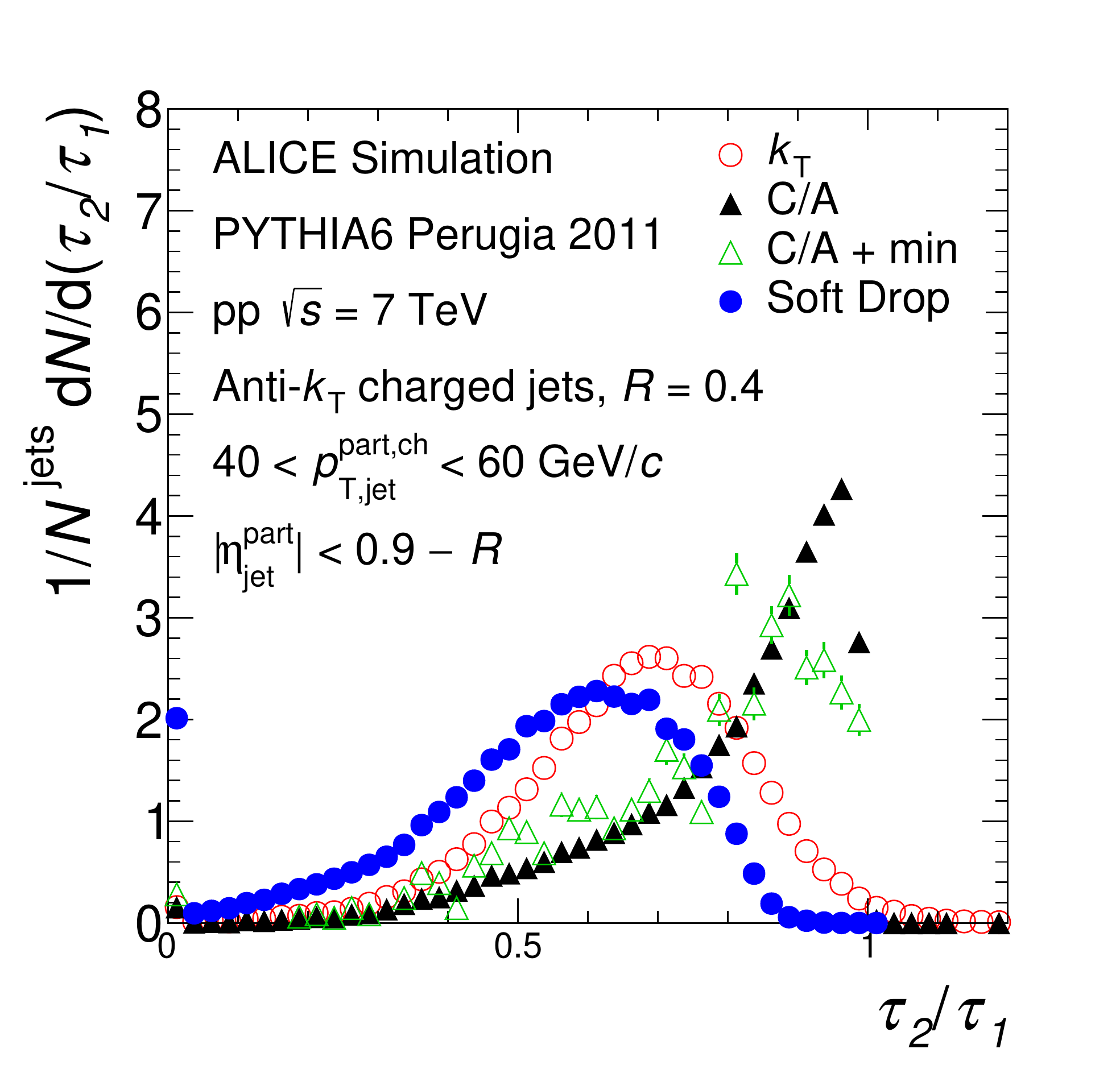}
\label{fig:pythiaAxes_Tau2to1}
\end{minipage}
\begin{minipage}[b]{0.49\textwidth}
\centering
\includegraphics[width=\textwidth]{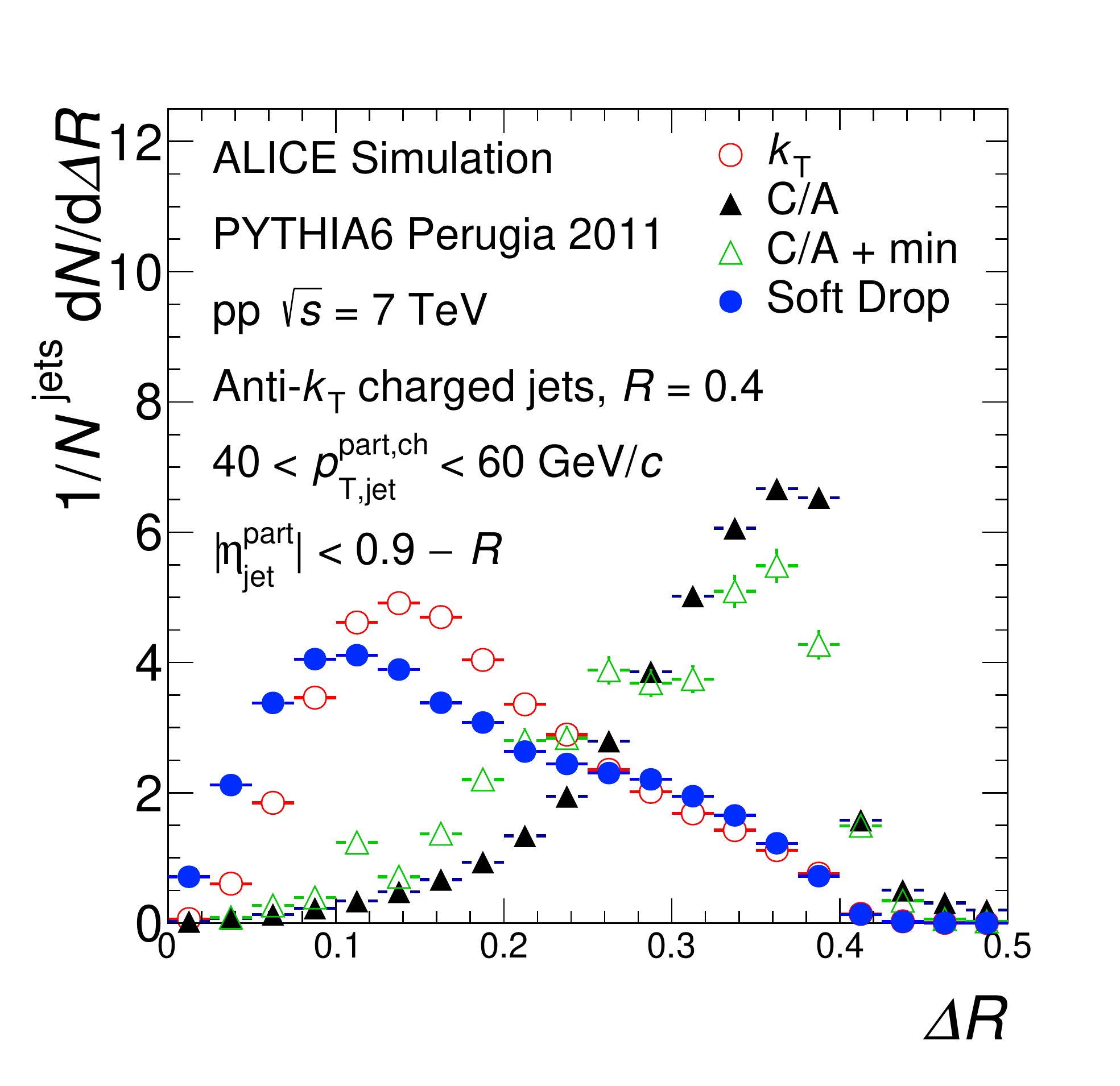}
\centering
\label{fig:pythiaAxes_DeltaR}
\end{minipage}
\caption{The $\tau_{2}/\tau_{1}$ (left) and $\Delta R$ (right) observables are shown for a variety of different axis selection algorithms. Both plots are obtained using a PYTHIA6 Perugia 2011 simulation of pp collisions at $\sqrt{s}=7$ TeV, in the particle level charged jet transverse momentum interval of $40 \leq p_{\rm T,\rm jet}^{\rm{part},\rm{ch}} \leq 60$ GeV/$c$.}
\label{fig:pythiaAxes}
\end{figure}

The comparison of the $\tau_{2}/\tau_{1}$ observable for different subjet axes choices, using a variety of reclustering algorithms, is shown in Fig.~\ref{fig:pythiaAxes} for track-based jets at particle level. These are calculated in simulated pp collisions at $\sqrt{s}=7$ TeV using the PYTHIA6 Perugia 2011 tune. The angular separation (in the $\eta$-$\varphi$ plane) of the subjet axes in the $\tau_{2}$ case, $\Delta R$, is also shown in Fig.~\ref{fig:pythiaAxes} (right). The $\tau_{2}/\tau_{1}$ distribution measured for the C/A case tends towards unity. This is due to the fact that C/A selects soft large-angle subleading prongs, to which the rest of the jet emissions are not aligned. The addition of the Soft Drop groomer significantly changes the subjet axes, with the $\tau_{2}/\tau_{1}$ distribution in this case peaking at lower values. This indicates that the subleading axis is correlated to a larger fraction of the $p_\mathrm{T,jet}$. There is also an increase in the population of jets with a value of $\tau_{2}/\tau_{1} = 0$, due to an increase of two-track jets after grooming. The distribution obtained with the Soft Drop groomer is qualitatively similar to the $k_\mathrm{T}$ case, where by construction soft particles are clustered together first and the final reclustering step brings together any hard structures present in the jet. 

In the ``C/A + min" case shown in the plots, a minimisation step is performed, starting with the axes returned by the C/A reclustering. The process seeks to find a local minimum of $N$-subjettiness, by geometrically varying the axes~\cite{Thaler:2011gf}. The minimisation was expected to reduce the sensitivity of the axes to uncorrelated soft radiation. However the impact of this additional step was found to be minimal, with respect to the C/A case without minimisation, both in the PYTHIA calculations presented here and in measurements made in pp and Pb--Pb collisions. As such, this particular configuration will not be discussed further.

The $\Delta R$ distributions, obtained using various reclustering algorithms, show that soft subleading prongs, selected by the C/A algorithm, are present at large angles from the jet core ($\Delta R > 0.3$). In contrast, hard subleading prongs, selected by both the $k_\mathrm{T}$ and Soft Drop groomer, appear at small angles ($\Delta R \sim 0.1$). This indicates that this sample of QCD jets are predominantly single-cored, with most of the $p_{\rm T,\rm jet}^{\rm ch}$ appearing close to the jet axis.

\begin{figure}[ht]
\centering
\begin{minipage}[b]{0.49\textwidth}
\centering
\includegraphics[width=\textwidth]{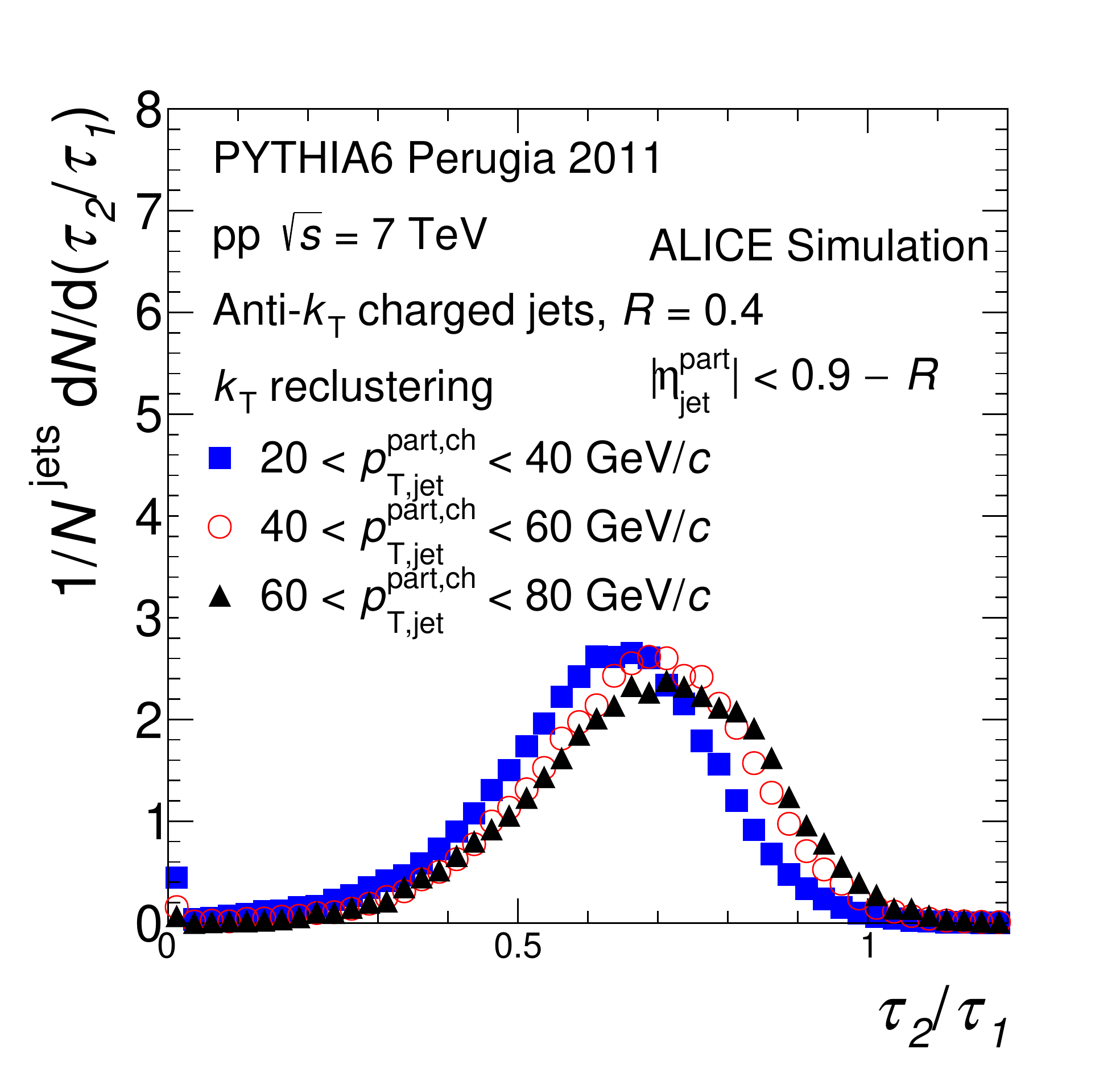}

\label{fig:pythiaPt_Tau2to1}
\end{minipage}
\begin{minipage}[b]{0.49\textwidth}
\centering
\includegraphics[width=\textwidth]{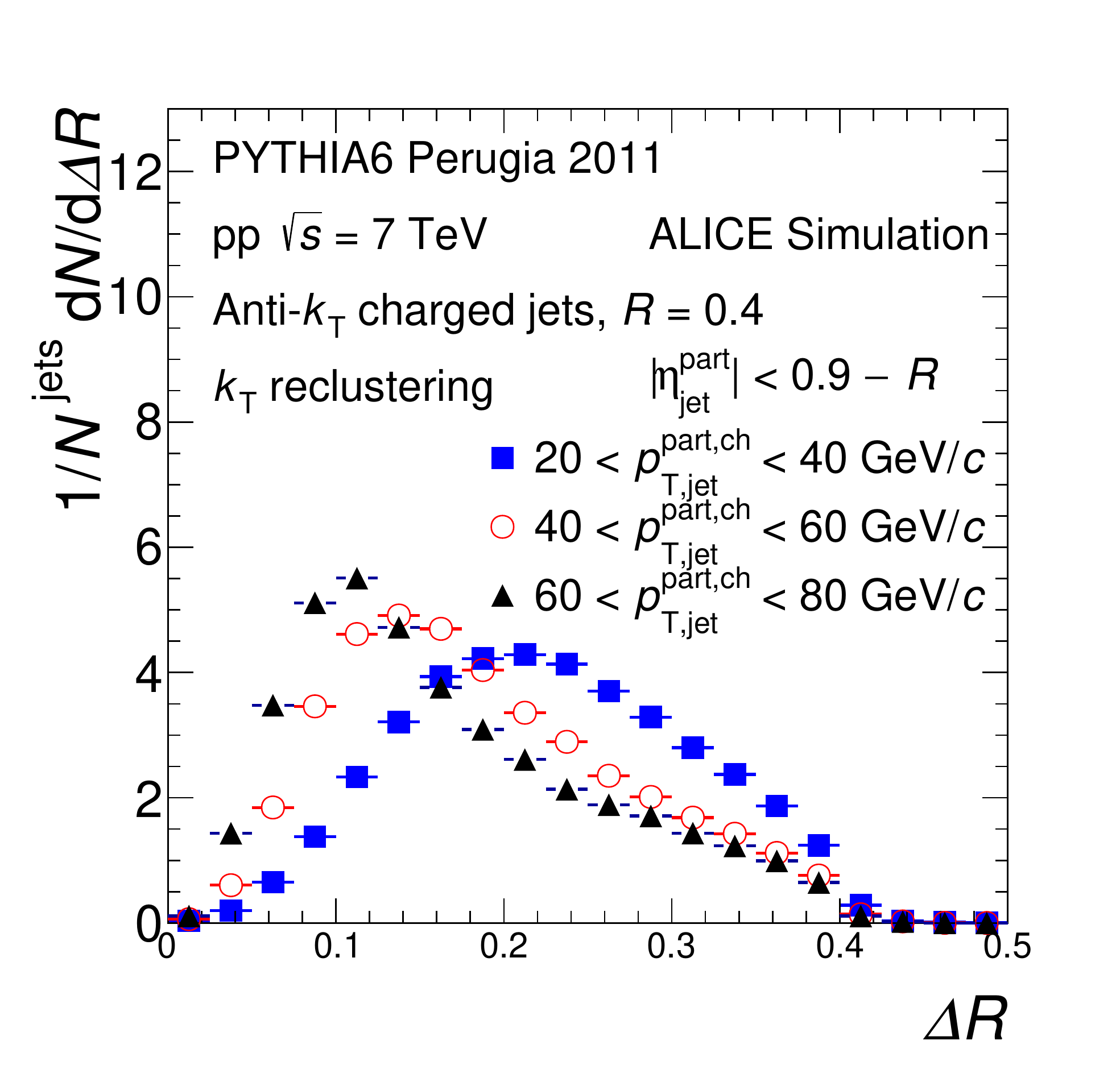}

\label{fig:pythiaPt_DeltaR}
\end{minipage}
\caption{The $p_{\rm T,\rm jet}^{\rm ch}$ dependence of the $\tau_{2}/\tau_{1}$ (left) and $\Delta R$ (right) observables are shown, for the axes obtained using the $k_{\rm{T}}$ reclustering algorithm. These calculations are made using PYTHIA6 Perugia 2011 simulations of pp collisions at $\sqrt{s}=7$ TeV.}
\label{fig:pythiaPtDependence}
\end{figure}

The $p_{\rm T,\rm jet}^{\rm ch}$ dependence of the $\tau_{2}/\tau_{1}$ and $\Delta R$ observables is shown in Fig.~\ref{fig:pythiaPtDependence} for the $k_{\rm{T}}$ reclustering algorithm, for three consecutive $p_{\rm T,\rm jet}^{\rm ch}$ intervals. The $\Delta R$ shape exhibits a strong $p_{\rm T,\rm jet}^{\rm ch}$ dependence in accordance with the collimation of jets with increasing $p_{\rm T,\rm jet}^{\rm ch}$. However, the $\tau_{2}/\tau_{1}$ shape shows only a mild dependence on the $p_{\rm T,\rm jet}^{\rm ch}$. The study was performed for the other algorithms (not shown), with the same conclusion also holding for C/A with Soft Drop grooming. However, no $p_{\rm T,\rm jet}^{\rm ch}$ dependence is observed for the C/A reclustering algorithm, with and without minimisation of the $\tau_{2}$ and $\tau_{1}$ variables. This is because the presence of large-angle soft particles in the jet cone is largely independent of $p_{\rm T,\rm jet}^{\rm ch}$. The weak $p_{\rm T,\rm jet}^{\rm ch}$ dependence observed for the $\tau_{2}/\tau_{1}$ observable facilitates a cleaner comparison of measurements made in the same $p_{\rm T,\rm jet}^{\rm ch}$ interval in Pb--Pb and pp collisions, as the reduction in the $p_{\rm T,\rm jet}^{\rm ch}$ scale due to quenching in Pb--Pb collisions has no effect on the vacuum properties of this observable.

%============================================================
\section{Data sets, event selection and simulations}
\label{sect:Dataset}
A detailed description of the ALICE detector and its performance can be found in Refs.\cite{Aamodt:2008zz,Abelev:2014ffa}. The analysed pp data were collected during Run 1 of the LHC in 2010 with a collision centre-of-mass energy of $\sqrts~= 7$ TeV using a minimum bias (MB) trigger. The MB trigger configuration is the same as described in Ref.\cite{ALICE:2014dla}.
The data from heavy-ion collisions were recorded in 2011 at $\sqrtsNN~= 2.76$ TeV. This analysis uses the $0$--$10\%$ most-central Pb--Pb collisions selected by the online trigger based on the hit multiplicity measured in the forward V0 detectors\cite{Abbas:2013taa}.
The data sets and event selections are identical to Refs.\cite{Adam:2015doa,Acharya:2018uvf}. After offline selection, the pp sample consists of 168 million events ($L_{\rm{int}} \approx 2.5\, \mathrm{nb}^{-1}$), while the Pb--Pb sample consists of 19 million events ($L_{\rm{int}} \approx 21.4\, \mathrm{\mu b}^{-1}$). In heavy-ion collisions a semi-inclusive hadron-jet procedure\cite{Adam:2015doa} is employed, as described in section~\ref{sect:JetReco}, which further restricts the event selection.

 The analysis uses primary charged particles reconstructed as tracks in the Inner Tracking System (ITS)~\cite{Aamodt:2010aa} and Time Projection Chamber (TPC)~\cite{Alme:2010ke} which both cover the full azimuth and pseudorapidity $|\eta| < 0.9$.  Tracks are required to have transverse momentum $0.15 <p_{\rm T} <100$ GeV/$c$. Further details of the track selection can be found in Ref.\cite{Adam:2015ewa}.  

In pp collisions, the tracking efficiency is approximately 80\% for tracks with $\pT > 1~\gev$, decreasing to roughly 56\% at $\pT = 0.15~\gev$, with a track momentum resolution of 1\% for $\pT = 1~\gev$ and 4.1\% for $\pT = 40~\gev$~\cite{Abelev:2014ffa,Abelev:2012hxa}. 
In $0$--$10\%$ most-central Pb--Pb collisions, the tracking efficiency is about 2 to 3$\%$ lower than in pp at any given $p_{\rm T}$. The track $p_{\rm T}$ resolution is about 1\% at $\pT = 1~\gev$ and 2.5\% for $\pT = 40~\gev$~\cite{Adam:2015ewa}.

For both centre-of-mass energies of $\sqrts~= 2.76$ TeV and $\sqrts~= 7$ TeV used in this analysis, pp collisions were simulated using PYTHIA6 (Perugia Tune 2011)~\cite{Skands:2010ak}. These Monte Carlo (MC) simulations were used to construct the response matrices employed as part of the unfolding procedures detailed in Section~\ref{sect:Unfolding}. These MC simulations were utilised at three levels which will be discussed in this paper. The first is the particle level MC, which includes primary particles and the decay products from strong and electromagnetic decays. This is used to construct the truth level axes of the response matrices. The second is the reconstructed level MC, which is obtained by propagating the particle level events through a detailed reconstruction of the ALICE apparatus using GEANT3~\cite{GEANT3}. This level includes both secondaries from interactions in the detector material and the products of weak decays and is used as the measured axes of the response matrix in pp collisions. To account for the smearing due to the background in Pb--Pb events, the MC simulated at $\sqrts~=2.76$ TeV is embedded into real, $0$--$10\%$ most-central, Pb--Pb events at reconstructed level to obtain the measured axes for the response matrices in this collision system. This comprises the third type of MC in use, referred to as the embedded level. PYTHIA-generated events are embedded instead of real pp data measured at $\sqrts~=2.76$ TeV, due to the limited size of the measured data sample. The differences between PYTHIA and measurements in pp collisions are studied at $\sqrts~=7$ TeV and are considered when comparing with measurements in Pb--Pb events.

\section{Jet reconstruction and underlying event corrections}
\label{sect:JetReco}

In both collisions systems jet reconstruction is performed on all accepted tracks, using the anti-$k_{\rm{T}}$ algorithm with a jet resolution parameter of $R=0.4$. This algorithm is the standard choice for jet finding at the LHC due to both the stable shape of the resulting jets as well as the disposition of the algorithm to cluster around hard structures. The choice of jet clustering algorithm determines the set of jet constituents, with no impact on the reclustering procedure beyond that. The $E$-scheme, which simply combines the four-vectors of two tracks, is used to recombine tracks~\cite{Cacciari:2011ma}, with the mass of each track assumed to be that of the charged pion. Jet finding is done using the FastJet package~\cite{Cacciari:2011ma}. The jet area, $A_{\rm jet}$, is calculated with a precision in the order of a percent, using so called ghost particle areas of $A_{\rm g} = 0.005$~\cite{FastJetArea}. In Pb--Pb collisions, a selection on the jet area is applied, such that jet candidates with an area smaller than 60$\%$ of the area of a cone with a radius of $0.4$ are rejected. This requirement suppresses background jets whilst
preserving true hard jets with a high efficiency~\cite{Jacobs:2010wq} (100$\%$ in the reported \pTjetch $\,$interval). Additional geometric selections are also imposed on the jet candidates, requiring $|\eta_{\rm{jet}}|<0.5$, where $\eta_{\rm{jet}}$ is the pseudorapidity of the jet axis. This ensures that the full jet cone is contained within the acceptance of the ALICE inner barrel.
The jet energy instrumental resolution is similar for pp  and Pb--Pb collisions, varying from 15\% at \pTjetch~= 20 \gev~to 25\% at \pTjetch~= 100 \gev. The Jet Energy Scale uncertainty is dominated by the tracking efficiency uncertainty which is 4$\%$.

In pp collisions, for the presented jet resolution parameter $R=0.4$ and jet momentum interval $\,\,\,\,\,\,\,\,\,\,\,\,\,\,\,\,\,\,\,\,\,\,\,\,\,\,\,\,\,\,$ $40<p_{\rm T,\rm jet}^{\rm ch}<60$ GeV/$c$, the impact of the underlying event background is minimal. This, in addition to the low pile-up contamination, means that underlying event corrections in this collision system are not required. However, for the Pb--Pb analysis, the presence of the large background due to the underlying event necessitates corrections both for the smearing of the jet $p_{\rm T}$ and jet substructure observables, as well as to the measured yield of jets. The smearing of the measured quantities for each jet is corrected for by subtracting the average underlying background in each event from each jet individually, as described in section~\ref{sect:AvBackg}. Any residual smearing due to local fluctuations of the background are corrected for using a two-dimensional bayesian unfolding procedure, as described in section~\ref{sect:Unfolding}.

The need for a correction to the measured yield of jets arises from a contamination by jets purely comprised of soft particles not correlated with a given hard scattering, known as combinatorial jets. This contamination must be corrected for before the unfolding procedure. Combinatorial jets constitute one of the main challenges for measurements at large jet resolution parameter and low $p_{\rm T,\rm jet}^{\rm ch}$ in heavy-ion collisions, as the rate of combinatorial jets is high in this regime. In order to correct for the combinatorial yield without introducing a bias on the jet fragmentation, a data driven semi-inclusive approach using hadron-jet coincidences is extended for the first time to a substructure measurement. This procedure is described in detail in section~\ref{sect:fakes}. The average underlying event is first removed on a jet-by-jet basis, before correcting the resulting yield of jets. Once the impact of combinatorial jets on the measured sample is suppressed, residual fluctuations of the underlying event are corrected for via an unfolding procedure.

\subsection{Average background subtraction in heavy-ion collisions}
\label{sect:AvBackg}
The constituent subtraction~\cite{Berta:2014eza} and area-derivatives~\cite{Soyez:2012hv} methods are applied independently to correct the candidate jet $p_{\rm T}$ and shape distributions for the average underlying event background, with the former used as the default method and the latter used as a systematic variation. To estimate the background in each event, a separate jet reconstruction pass is carried out using the \kT\ algorithm with $R = 0.2$. This algorithm is sensitive to soft clusters in the event and allows for the estimation of the density of jet-like transverse momentum and mass due to the background, $\rho$ and $\rho_{\rm m}$. These are defined as
\begin{equation}
\rho=\mathrm{median}\left( \frac{p_{\rm {T,jet}}^{\rm{raw},\rm{ch},\it{i}}}{A_{\rm{jet}}^{\it{i}}}\right),\, \, \, \, \,\rho_{m}=\mathrm{median}\left( \frac{\it{m}_{\rm{jet}}^{\rm{raw},\it{i}}}{A_{\rm jet}^{\it{i}}}\right),
\label{eq:rho}
\end{equation}

\noindent
where the index $i$ runs over all jet candidates in an event, excluding the two with the highest $p_{\rm T,\rm jet}^{\rm raw, \rm ch, \rm i}$. The transverse momentum, mass and area of the
$i^{\textup{th}}$ reconstructed jet are denoted by  $p_{\rm T,\rm jet}^{\rm raw,\rm ch,\rm i}$, $m_{\rm{jet}}^{\rm{raw},\it{i}}$
and $A_{\rm jet}^{i}$, respectively. 

The constituent subtraction method works by uniformly adding ghost particles to each jet, with the $p_{\rm T}^{\rm ch}$ and mass of these particles being scaled to represent the event-wise $\rho$ and $\rho_{\rm m}$ values, respectively. Jet constituents and neighbouring ``ghosts" are iteratively paired up and their $p_{\rm T}^{\rm ch}$ and mass are subtracted. The pairing and subtraction are done separately for the $p_{\rm T,\rm jet}^{\rm ch}$ and mass. During this subtraction the $p_{\rm T}^{\rm ch}$ (or mass) of the jet constituent and ``ghost" in each pair are compared and the smaller $p_{\rm T}^{\rm ch}$ (or mass) value is subtracted from the larger one. The $p_{\rm T}^{\rm ch}$ (or mass) of the smaller one is then set to zero and the jet constituent or ``ghost" that it belonged to is removed from the next iteration of pairing. In this way, the average underlying event background is removed at a constituent level from the jets. The area-derivatives method employs a numerical approach to account for the underlying background. The jet is populated with ``ghost" particles and the background subtracted observable of interest is redefined as an expansion containing the observable measured in the presence of the background and its derivatives (in this work up to the second order) with respect to the $p_{\rm T}^{\rm ch}$ scale of the ``ghosts", which are subtracted from the first term. This series is then numerically solved in the limit that the ``ghost" $p_{\rm T}^{\rm ch}$ scale goes to zero, which represents the case with no underlying event background. A detailed description of the methods and their application to jet substructure observables in ALICE can be found in~\cite{Acharya:2018uvf}.

\subsection{Suppression of combinatorial jets via hadron-jet correlations in heavy-ion collisions}
\label{sect:fakes}
In order to suppress the combinatorial jet yield in the measured sample, a data-driven method using semi-inclusive hadron-jet correlations, is applied. First, two exclusive classes of high trigger charged hadron $p_{\rm T}$, with $15 \leq p_{\rm T} < 45$ GeV/{\it{c}} (Signal) and $8 \leq p_{\rm T} < 9$ GeV/$\it{c}$ (Reference), denoted as TT$\{15,45\}$ and TT$\{8,9\}$ (where TT stands for trigger track), are considered. These intervals are similar to those used in\cite{Adam:2015doa} and are chosen so as to optimise the signal-to-background ratio, as described below. The event sample is split into two statistically independent samples for the signal and reference measurements. Jet finding is then performed in events containing a trigger hadron, with the jets constrained to a back-to-back (with respect to the trigger hadron) azimuthal window, $\Delta \varphi = \varphi_{\rm{TT}} - \varphi_{\rm{jet}}$, which is defined such that $\lvert \pi$-$\Delta \varphi \rvert <0.6$. This region of azimuthal phase space is chosen to account for in-medium deflections of the recoiling parton. The same pseudorapidity window of $|\eta| < 0.5$, as described in Section~\ref{sect:JetReco}, is used for jet selection. The contribution of combinatorial jets found in the  recoiling region of the high transverse momentum trigger hadrons is expected to be uncorrelated to these hadrons and hence equal for both TT classes. Therefore, subtracting the per-trigger normalised yield of jets measured in the recoil regions of the two classes results in a combinatorial-suppressed distribution which allows for unfolding to particle level. The two trigger classes are of sufficiently high $p_{\rm T}$ so that topological, multiplicity and flow biases induced by a trigger hadron saturate and are removed by the subtraction procedure. In this paper the hadron-jet recoil method is applied for the first time to a jet substructure measurement, where the subtraction involves the two-dimensional recoil yields of $p_{\rm T,\rm jet}^{\rm ch}$ and the $\tau_{2}/\tau_{1}$ substructure observable (defined in Section~\ref{sect:definition}) measured for each trigger hadron class. The subtraction procedure is given by,

\begin{equation}
\Delta_\mathrm{recoil}^{\tau_{2}/\tau_{1}}=
\frac{1}{N_{\rm{trig,Sig}}} \frac{\rm{d}^2\it{N}}{\rm{d}\it{p}_{\rm{T,jet}}^{\rm ch}\rm{d}\tau_{2}/\tau_{1}}\Bigg\vert_{\pTtrig\in{\TTSig}} -
\frac{1}{N_{\rm{trig, Ref}}} \frac{\rm{d}^2\it{N}}{\rm{d}\it{p}_{\rm{T,jet}}^{\rm ch}\rm{d}\tau_{2}/\tau_{1}}\Bigg\vert_{\pTtrig\in{\TTRef}},
\label{eq:DRecoilShape}
\end{equation}

where \TTSig $\,$ and \TTRef $\,$ represent TT$\{15,45\}$ and TT$\{8,9\}$, respectively. The variables $N_{\rm{trig,Sig}}$ and $N_{\rm{trig,Ref}}$ represent the number of trigger hadrons selected in the signal and reference classes, respectively. It should be noted that the scale factor term found in\cite{Adam:2015doa}, which corrects the reference trigger class yield for effects of finite phase-space in the recoil region, is omitted. The impact of this scale factor is found to be negligible through tests in data. This follows from the fact that this correction mostly affects the absolute yield of the subtracted distribution and has a smaller impact on the shape of the distribution. As the PYTHIA distributions are compared with inclusive measurements in pp collisions, only the descriptions of the shapes of the distributions are validated. As the descriptions in PYTHIA of the magnitudes of the semi-inclusive yields of the observables are not validated,  the presented measurements in Pb--Pb collisions are self normalised for comparisons to PYTHIA. As such, the scale factor correction has a negligible impact on the results.

\subsection{Performance of the heavy-ion underlying event subtraction procedures on $\tau_{2}/\tau_{1}$ }

\begin{figure}[h!]
\centering
\begin{minipage}[b]{0.7\textwidth}
\centering
\includegraphics[width=\textwidth]{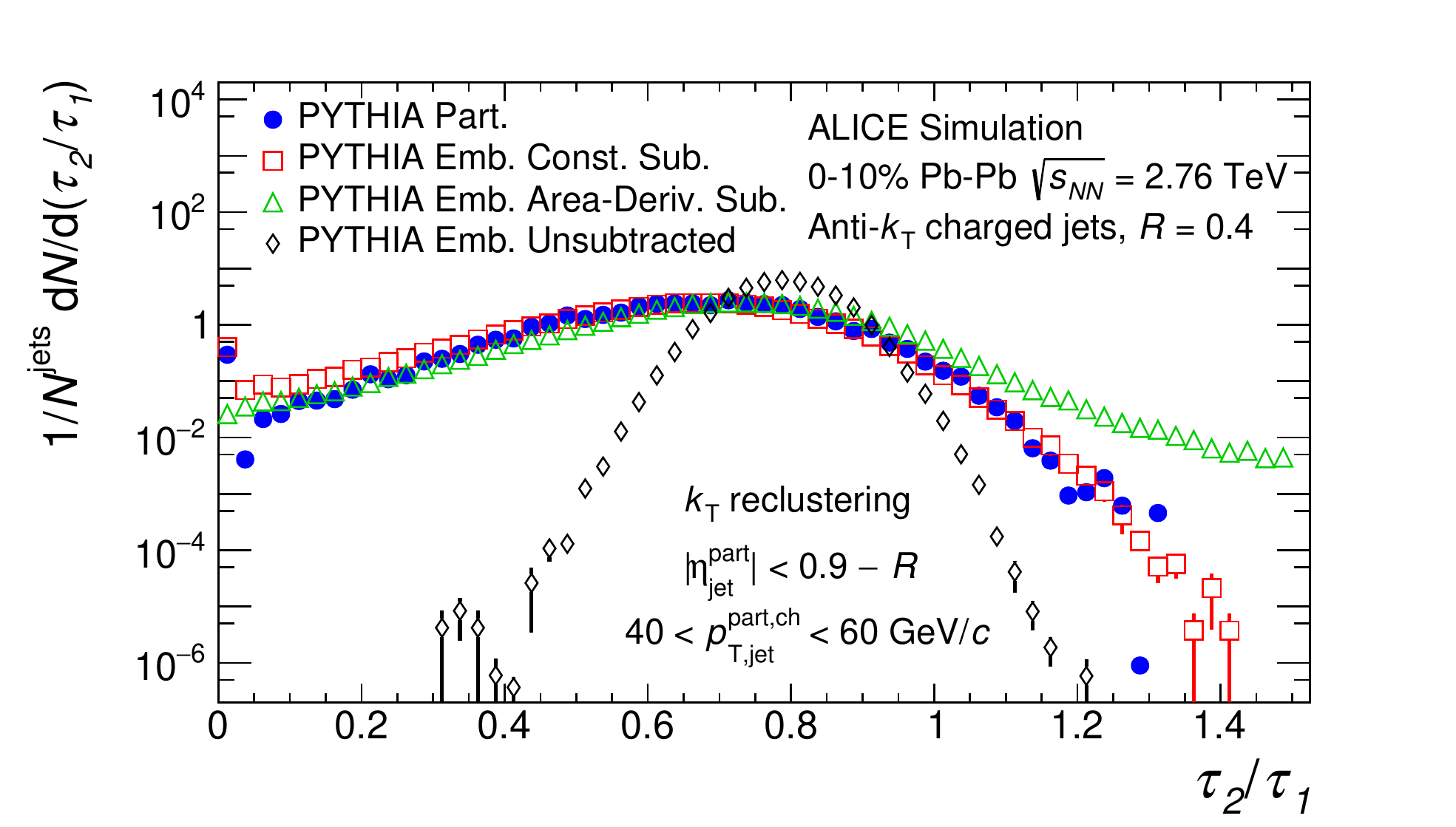}
\end{minipage}
\centering
\begin{minipage}[b]{0.7\textwidth}
\centering
\includegraphics[width=\textwidth]{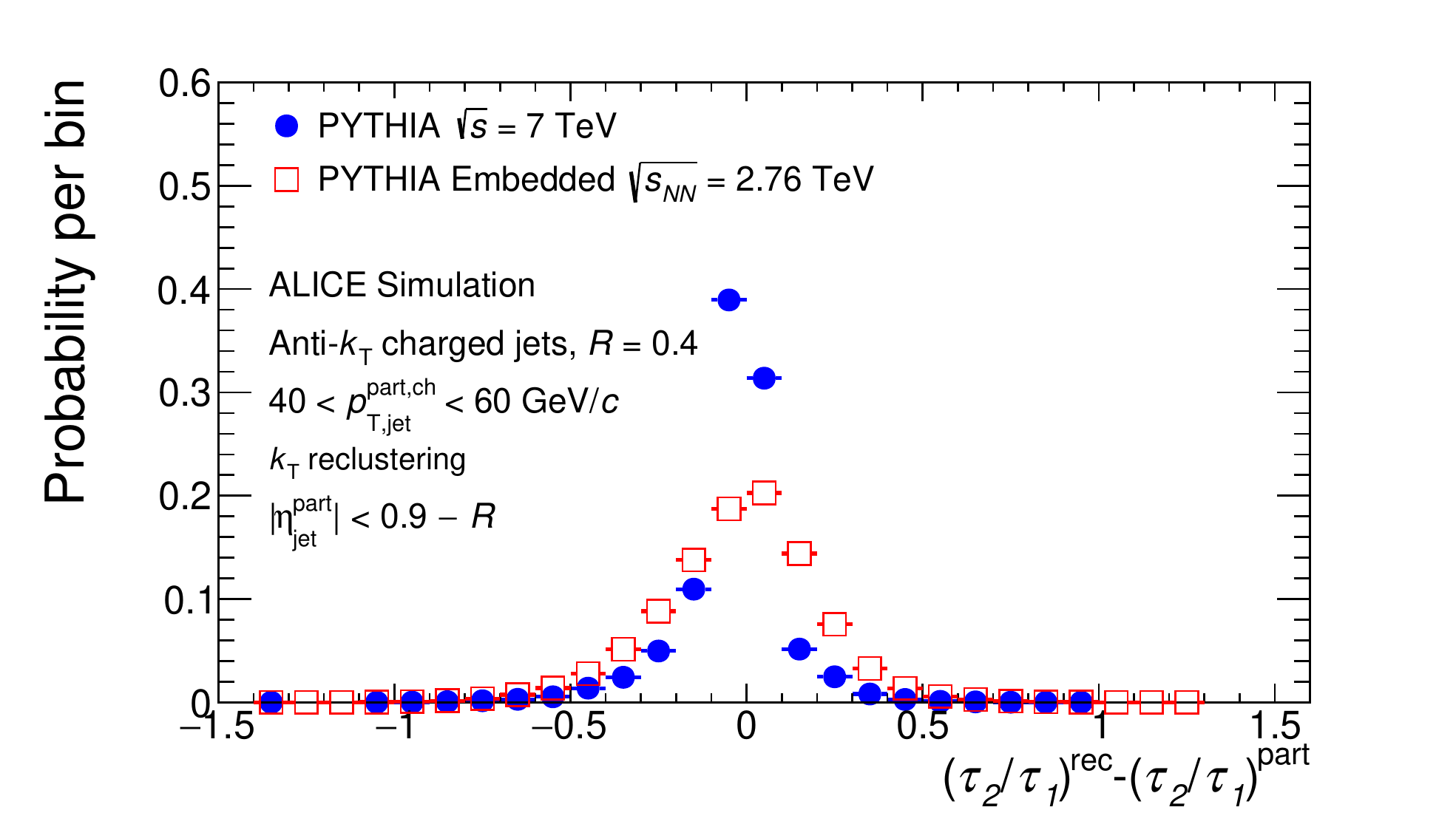}
\end{minipage}
\caption{Left: The performances of the constituent (squares) and area-derivatives (triangles) subtraction methods on the $\tau_{2}/\tau_{1}$ observable, measured with the $k_\mathrm{T}$ reclustering algorithm, are evaluated by embedding PYTHIA MC events into real Pb--Pb data. The filled circles represent the case with no heavy-ion background and the diamonds represent the case with background but no subtraction. Right: The $\tau_{2}/\tau_{1}$ residual distributions characterising the observable resolution due to detector effects and uncorrelated background. Both plots are shown for $40 < p_{\rm T,\rm jet}^{\rm part,\rm ch} < 60$ GeV/$\it{c}$.}
\label{fig:BackSubPerform}
\end{figure}

The performances of the constituent subtraction and area-derivatives methods for the $\tau_{2}/\tau_{1}$ observable are studied by embedding simulated PYTHIA MB events, propagated to reconstructed level, into the $0$--$10\%$ most central, real Pb--Pb events. Jet finding is then performed at the embedded level and the constituent subtraction and area-derivatives methods are applied. The particle level jets are then matched to their embedded level counterparts to assess the impact of these underlying event subtraction methods. In addition to determining a geometric correspondence between the particle level and embedded level jet axes, the matching condition also requires that tracks pertaining to half of the transverse momentum of the particle level jet are included in the matched embedded level jet. These will be the matching criteria used in all cases between jets at particle and embedded levels.  Figure~\ref{fig:BackSubPerform} (left) shows the $\tau_{2}/\tau_{1}$ distributions, measured using the $k_{\rm{T}}$ reclustering algorithm, for unsubtracted embedded  jets (diamonds), average background-subtracted embedded jets (squares and triangles for the two methods), and PYTHIA particle level jets (full circles). The heavy-ion background has a large impact on the PYTHIA distribution, promoting the unsubtracted embedded distribution to significantly larger values compared to particle level. The background subtraction techniques do a fairly good job of correcting the distribution back to particle level. The comparison is performed in the particle level momentum, $p_{\rm T,\rm jet}^{\rm part,\rm ch}$, interval of $40$--$60$\,GeV/$c$. Similar performances are obtained for all subjet axes choices. After applying these methods to the measured data, any remaining residual differences between the particle level and background subtracted distributions, which are due to background fluctuations and detector effects, are corrected using an unfolding procedure (see Section~\ref{sect:Unfolding}). Since the embedded MC jets are matched to particle level, no yield correction to suppress combinatorial jets is required for this performance check.

Figure~\ref{fig:BackSubPerform} (right) shows the residual distributions for $\tau_{2}/\tau_{1}$, simulated for both pp and Pb--Pb (via embedding) collisions, for $40 < p_{\rm T,\rm jet}^{\rm part, \rm ch} < 60$ GeV/{\it{c}}. As expected, the distributions measured in Pb--Pb collisions exhibit a larger width compared to those in pp collisions, as the residuals are influenced by background fluctuations in addition to instrumental effects.

\begin{figure}[h!]
\centering
\begin{minipage}[b]{0.69\textwidth}
\centering
\includegraphics[width=\textwidth]{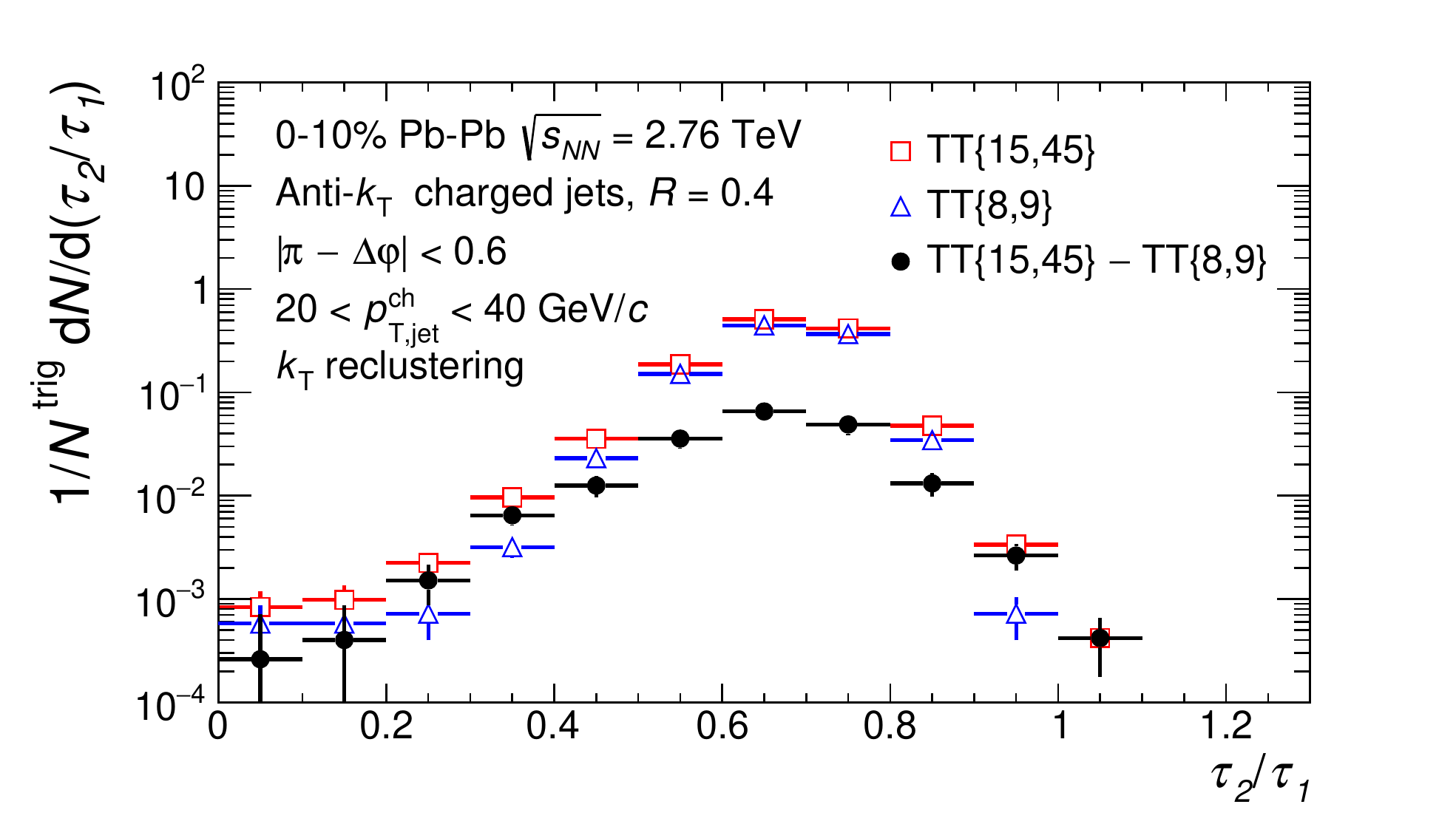}

\end{minipage}
\begin{minipage}[b]{0.69\textwidth}
\centering
\includegraphics[width=\textwidth]{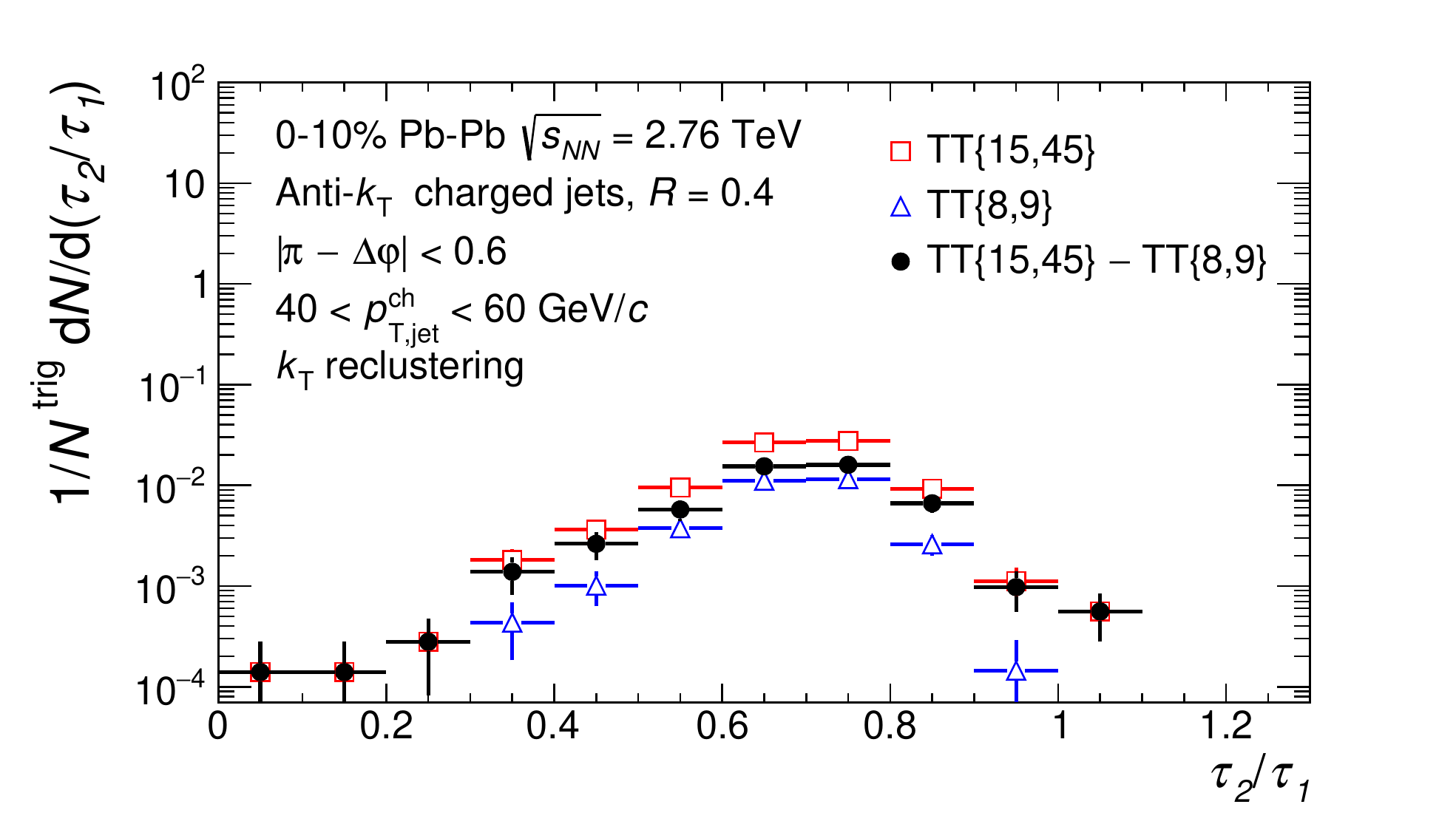}

\end{minipage}
\centering
\begin{minipage}[b]{0.69\textwidth}
\centering
\includegraphics[width=\textwidth]{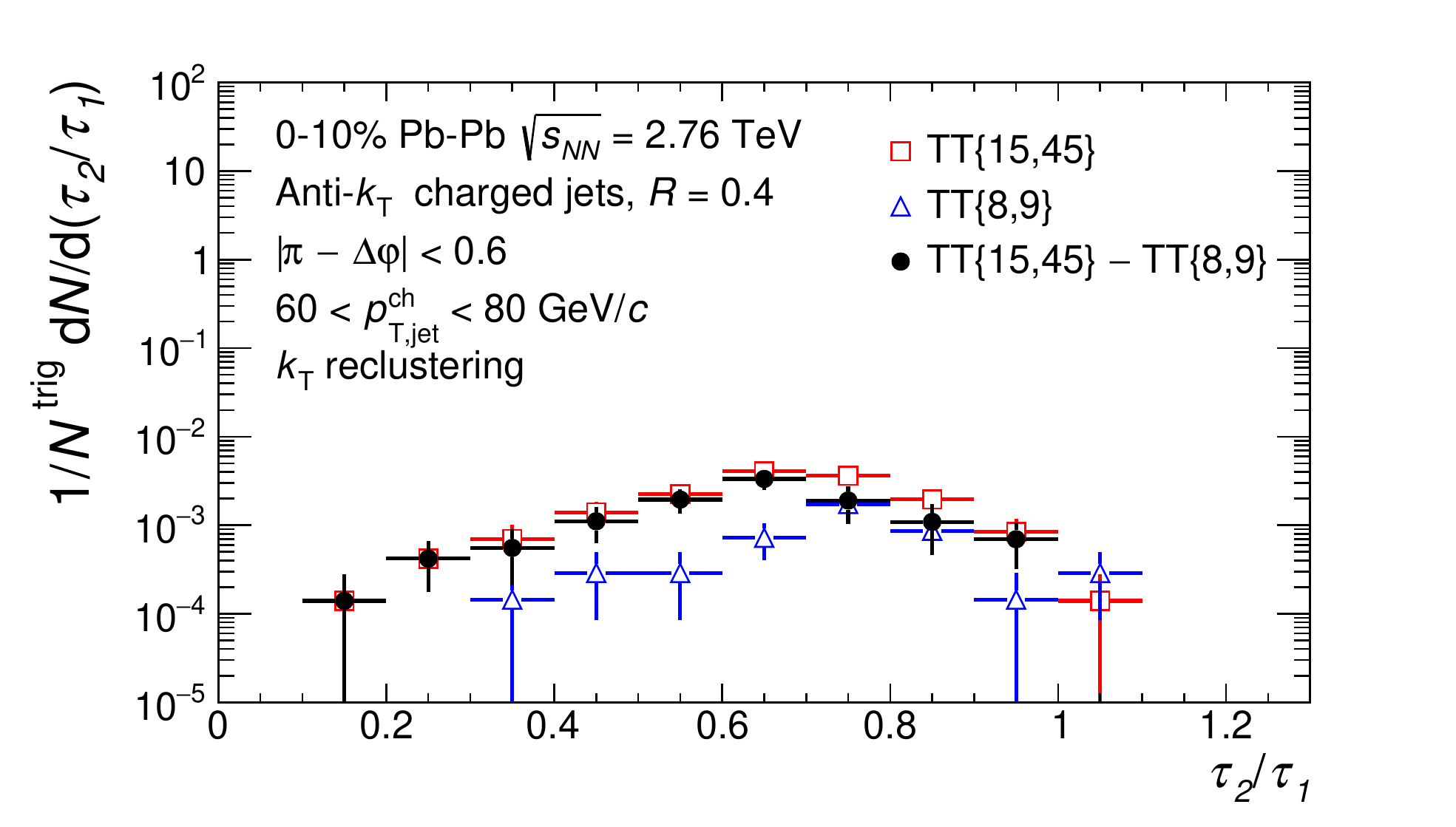}

\end{minipage}
\caption{The uncorrected $\tau_{2}/\tau_{1}$ distributions of the jets recoiling from a trigger hadron, measured in Pb--Pb collisions at $\sqrt{s_{\rm NN}} = 2.76$ TeV, are presented.  For each trigger hadron class, three consecutive uncorrected $p_{\rm{T},\rm{jet}}^{\rm{ch}}$ bins of $20 \leq p_{\rm{T},\rm{jet}}^{\rm{ch}} < 40$, $40 \leq p_{\rm{T},\rm{jet}}^{\rm{ch}} < 60$ and $60 \leq p_{\rm{T},\rm{jet}}^{\rm{ch}} < 80$ GeV/${\it c}$ are measured. The red, blue and black data points represent the signal trigger class, the reference trigger class and the difference between the two classes, respectively.}
\label{fig:HadronJetRaw}
\end{figure}

The impact of employing the semi-inclusive hadron-jet coincidence technique to suppress the combinatorial yield is shown for the measured data in Fig.~\ref{fig:HadronJetRaw}, where the $\tau_{2}/\tau_{1}$ distributions measured in the recoil region of each trigger hadron class, along with the difference of these two distributions, are presented in three intervals of $p_{\rm T,\rm jet}^{\rm ch}$. It can be seen that the signal-to-background ratio, which can be inferred from the separation of the two TT classes, increases with increasing $p_{\rm T,\rm jet}^{\rm ch}$, as the contribution of combinatorial jets decreases. For $20 < p_{\rm T,\rm jet}^{\rm ch} < 40$ GeV/$c$, combinatorial jets are dominant and the difference between the recoil shape yields of the two trigger classes is small. As expected, this difference, which is the combinatorial-suppressed distribution, increases relative to the TT$\{15,45\}$ distribution with increasing $p_{\rm T,\rm jet}^{\rm ch}$. It is now possible to unfold this distribution to particle level, as the combinatorial entries are heavily suppressed through this semi-inclusive correction.

%================================================

\section{Detector and background response and two-dimensional unfolding}
\label{sect:Unfolding}

\noindent The measured distributions are unfolded to simultaneously correct the reconstructed $p_{\rm T,\rm jet}^{\rm ch}$ and shape distributions back to particle level. The corrections account for detector effects (pp and Pb--Pb collisions) and residual background fluctuations (Pb--Pb collisions). As shown in Fig.~\ref{fig:BackSubPerform}, the background subtraction techniques applied in Pb--Pb collisions perform well for the measured $\tau_{2}/\tau_{1}$ distributions. Therefore, the unfolding procedure does not induce large correction factors.

A two-dimensional Bayesian unfolding procedure, as implemented in the RooUnfold package~\cite{RooUnfold}, is used. To guarantee statistical stability of the correction procedure, the 2D correlation ($p_{\rm T,\rm jet}^{\rm rec,\rm ch}$, shape$_{\rm jet}^{\rm{rec,ch}}$), which is the input to the unfolding, is binned such that there are no empty bins. This sets the upper limit of the input $p_{\rm T,\rm jet}^{\rm rec,\rm ch}$ range at $80$ GeV/$c$ in both pp and Pb--Pb collisions.
The response matrix is constructed with matched particle and embedded (reconstructed) level jets for the unfolding of Pb--Pb (pp) collisions and contains no entries for combinatorial jets. As the raw input data in Pb--Pb collisions also has a negligible contribution from combinatorial background jets, it can in principle be unfolded down to arbitrarily low jet $p_{\rm T,\rm jet}^{\rm rec,\rm ch}$. In practice, however, the minimum accepted jet momentum is set at 20 GeV/$c$ to maintain the statistical stability of the unfolding.  The same lower limit is also applied in pp collisions. The reconstructed range supplied to the response matrix must also mirror that of the input.

A particle level $p_{\rm T,\rm jet}^{\rm part, \rm ch}$ interval of 0--160\,GeV/$c$ was supplied to the response matrices.
The particle level ranges were extended relative to the input ranges, to account for jet migrations into the reconstructed level range due to
background fluctuations and instrumental effects.  
The unfolding procedure cannot account for the feed-in from reconstructed level jets outside
the truncated (reconstructed level) range which at particle level are within the given range. These must be corrected for independently. This correction is calculated using MC simulations at particle and reconstructed levels and is limited by
choosing reported unfolded bins far away from the truncation
thresholds. Therefore, our final results are presented for
the jet momentum interval $40$--$60$\,GeV/$c$.

For the Pb--Pb analysis, the entries into the response matrices are weighted according to their particle level $p_{\rm T,\rm jet}^{\rm part,\rm ch}$ values, to transform the shape of the particle level embedded spectrum from that of inclusive to recoil jets. This is to ensure the response matrix is more representative of the underlying input data distribution. These weights are calculated in bins of particle level $p_{\rm T,jet}^{\rm part,ch}$ by taking the ratio of a particle level PYTHIA recoil jet $p_{\rm T,\rm jet}^{\rm part,\rm ch}$ distribution and a (scaled) PYTHIA inclusive jet $p_{\rm T,\rm jet}^{\rm part,\rm ch}$ distribution.

The unfolded solutions converge after a few iterations in both collision systems. The stability of each unfolding procedure is tested by refolding the unfolded solutions back and checking the agreement with the input distributions. In Pb--Pb (pp) collisions, both distributions agree within 5$\%$ (1$\%$) after the third (second) iteration. A closure test is also performed, where two
statistically independent MC samples are used to fill the response matrix (particle and reconstructed levels) and
provide the input data to the unfolding (reconstructed level). In this case, the unfolded solutions agree with the particle level MC corresponding to the pseudo input samples within 10$\%$ in pp and Pb--Pb collisions.

\section{Systematic uncertainties}
\label{sect:sysuncert}
The systematic uncertainties for the jet substructure distributions are determined by varying the analysis settings for instrumental responses and method induced selections. The sources of systematic uncertainty considered are:
\begin{itemize}

\item The tracking efficiency uncertainty for the track selection used is $\pm 4\%$ in both pp and Pb--Pb collisions. The unfolding procedure is repeated with a response matrix constructed with $4\%$ worse tracking efficiency. This represents the largest contribution to the jet energy scale uncertainty~\cite{Acharya:2017goa}.

\item A series of systematic variations of the unfolding procedure are considered. The variation giving the largest uncertainty in each bin is chosen as the unfolding systematic uncertainty. These include the following:
\begin{enumerate}
\item The prior in the two-dimensional Bayesian implementation of RooUnfold is taken as the projection of the response matrix onto the particle level axis. The default prior is PYTHIA6 Perugia 0. A variation to the prior, which re-weights the response matrix such that the prior coincides with the unfolded solution, is considered.

\item  The unfolding is regularised by controlling the number of iterations performed, which is four for pp and Pb--Pb collisions in the default solutions. The uncertainty in the regularisation is estimated by considering the differences to solutions for iterations that are one lower and three higher than the default number of iterations.

\item In both collision systems, the minimum accepted jet $p_{\rm T,\rm jet}^{\rm ch}$ as input to the unfolding is 20\,GeV/$c$. As a variation, this truncation threshold is lowered by 10\,GeV/$c$. For Pb--Pb collisions, the $\tau_{2}/\tau_{1}$ shape input is also truncated at 1.0 to preserve statistical stability in the default unfolding case. The upper bound of this input is extended to 1.2 as a variation. 

\item The binning of the input data is varied arbitrarily (whilst keeping the statistical requirements of no empty bins) in both the $p_{\rm{ T,jet}}^{\rm ch}$ and shape dimensions.
\end{enumerate}

\item The choice of the background subtraction method in Pb--Pb collisions, which primarily affects the tails of the distributions, is varied. The default solution is obtained using the constituent subtraction method, whereas the systematic uncertainty is calculated using the area-based derivatives subtraction method. This uncertainty is not available for the distribution calculated using the Soft Drop algorithm as the numerical subtraction procedure does not produce stable results.

\end{itemize}

\vspace*{30px}

The different components of the systematic uncertainties, for the observables calculated using the $k_{\rm{T}}$ reclustering algorithm, are summarised in Tables~\ref{tab:Systpp}~and~\ref{tab:SystPbPb}, for pp and Pb--Pb collisions,
respectively.  All sources of systematic uncertainty are considered to be uncorrelated and each is symmetrised before being added in quadrature to obtain the overall systematic uncertainty.

 The relative contribution and magnitude of the uncertainties are similar for all the reclustering algorithms considered, with the exception of the first two bins of the $\Delta R$ distribution measured with the C/A algorithm in pp collisions, where the tracking efficiency uncertainty has a larger impact.

\begin{table}[h]
\centering
\renewcommand{\arraystretch}{1.3}
\caption{Relative systematic uncertainties of the measured observables, using $k_{\rm{T}}$ reclustering, in pp collisions at $\sqrt{s}=7$ TeV are shown for three selected intervals of the observables in the jet $p_{\mathrm{T,jet}}^{\rm ch}$ range of $40$--$60$\,GeV/$c$. All sources of systematic uncertainty are considered to be uncorrelated.}
\begin{tabular}{|l|ccc|ccc|}
\hline 
\multicolumn{1}{|l|}{Observable} & \multicolumn{3}{c|}{$\tau_{2}/\tau_{1}$} 
 & \multicolumn{3}{c|}{$\Delta R$} \\
\hline
\hline
\multicolumn{1}{|l|}{Interval} 
& 0.2--0.4 & 0.5--0.6 & 0.8--0.9 & 0.0--0.1 & 0.1--0.2 & 0.2--0.3  \\
\cline{1-7}

Tracking              & $7.5\%$ & $5.2\%$ & $11.2\%$
                    
                     & $4.4\%$ & $0.8\%$  & $0.9\%$ \\

Unfolding            & $1.5\%$ & $4.4\%$  & $5.7\%$ 
                      
                       & $3.7\%$  & $2.8\%$  & $3.9\%$   \\

 \cline{1-7}
 Total  & $7.6\%$ & $6.8\%$  & $12.6\%$
                    
                      & $5.7\%$ & $2.9\%$ & $4.0\%$ \\
                     
\hline
\end{tabular}
\label{tab:Systpp}
\end{table}

\begin{table}[h]
\centering
\renewcommand{\arraystretch}{1.3}
\caption{Relative systematic uncertainties on the $\tau_{2}/\tau_{1}$ shape, measured using $k_{\rm{T}}$ reclustering, are shown in $0$--$10\%$ most central Pb--Pb collisions at $\sqrt{s_{\rm NN}}=2.76$ TeV for three selected intervals of the observables in the jet $p_{\rm{T,jet}}^{\rm ch}$ range $40$--$60$\,GeV/$\it{c}$. All sources of systematic uncertainty are considered to be uncorrelated.}
\label{tab:SystPbPb}
\begin{tabular}{|l|ccc|}
\hline 
\multicolumn{1}{|l|}{Shape} & \multicolumn{3}{c|}{$\tau_{2}/\tau_{1}$} \\
\hline
\hline
\multicolumn{1}{|l|}{Shape interval} 
& 0.2--0.4 & 0.4--0.6 & 0.8--1\\
\cline{1-4}

Tracking   & $3.9\%$ & $1.5\%$ & $1.4\%$  \\

Unfolding                 & $11.6\%$ & $2.0\%$ & $3.2\%$ \\

Bkg. Sub                  & $2.4\%$ & $2.6\%$ & $0.9\%$ \\

 \cline{1-4}
Total                   & $12.5\%$ & $4.1\%$  & $3.6\%$  \\
\hline
\end{tabular}
\end{table}

\section{Results and discussion}
\label{sect:results}

\begin{figure}[h!]
\centering
\begin{minipage}[b]{0.49\textwidth}
\centering
\includegraphics[width=\textwidth]{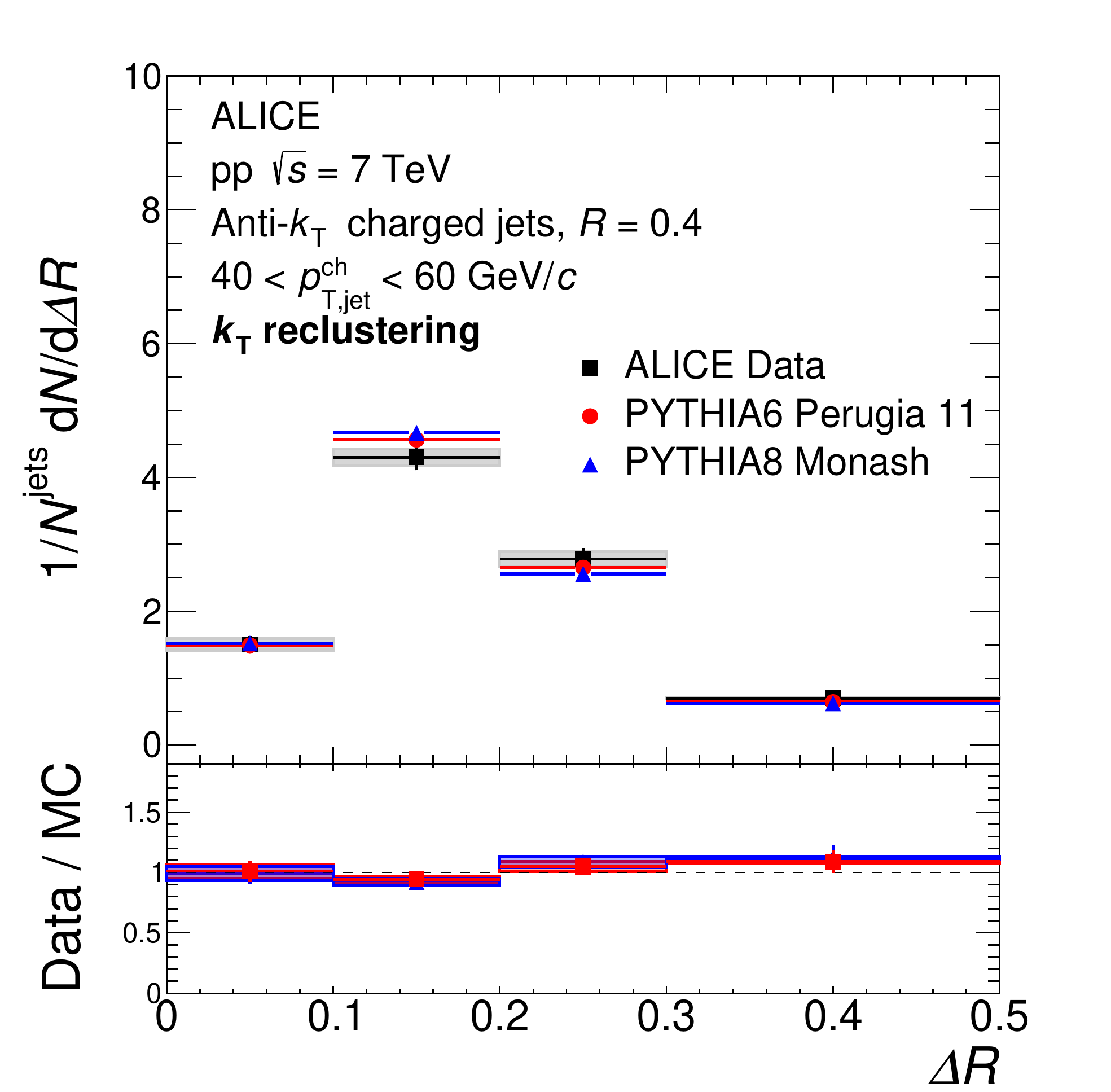}

\end{minipage}
\centering
\begin{minipage}[b]{0.49\textwidth}
\centering
\includegraphics[width=\textwidth]{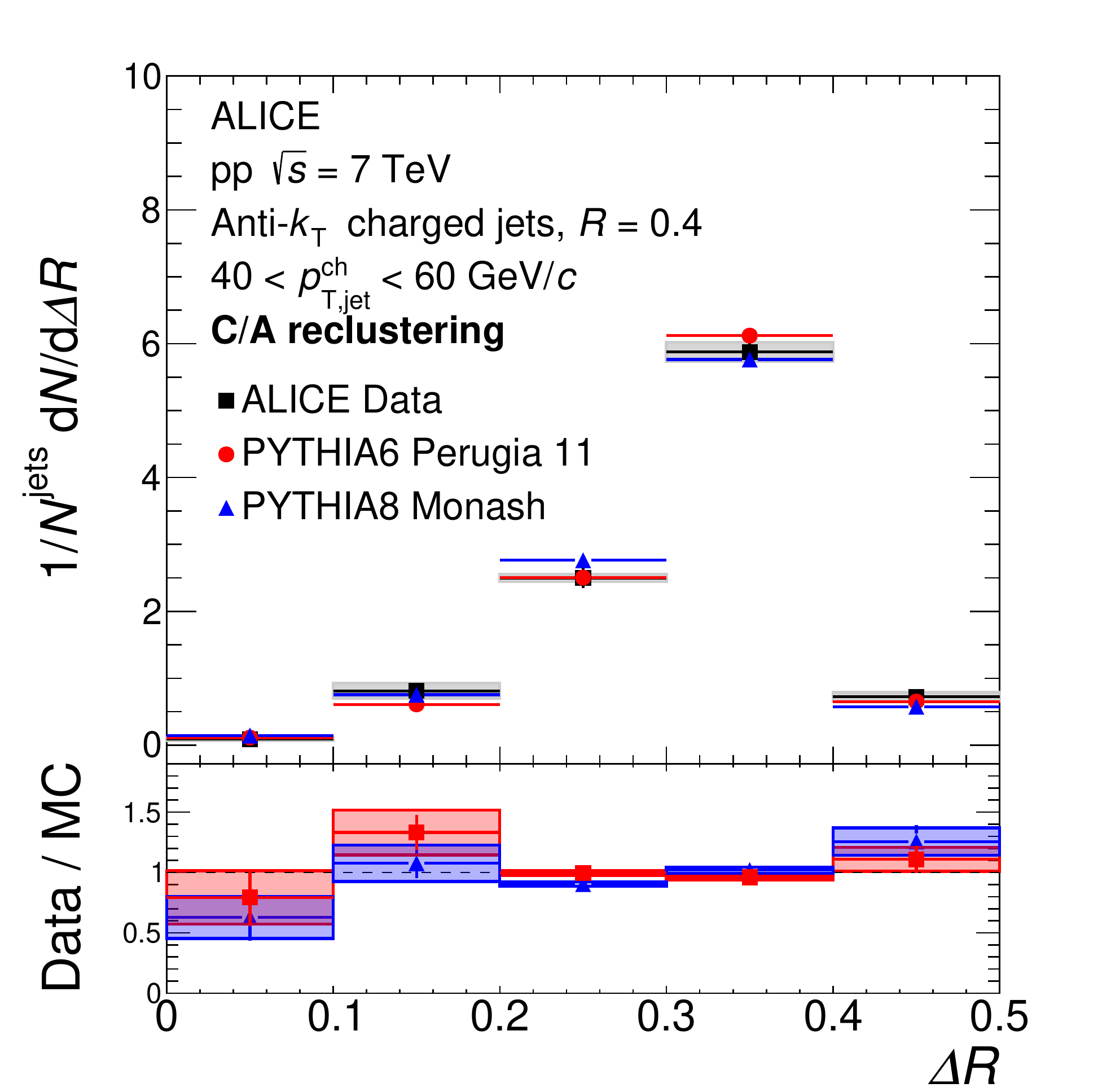}

\end{minipage}
\centering
\begin{minipage}[b]{0.49\textwidth}
\centering
\includegraphics[width=\textwidth]{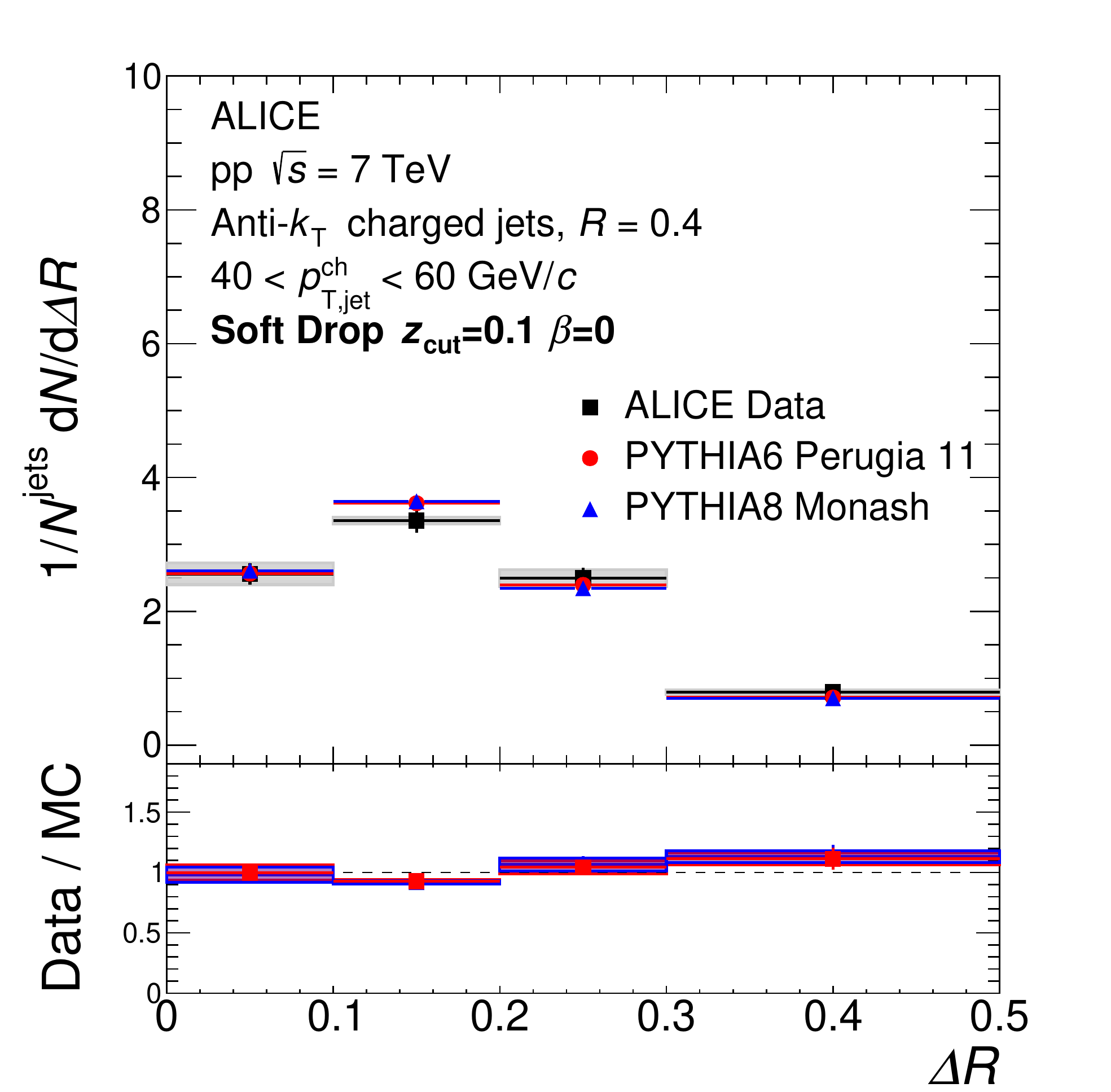}
\end{minipage}

\caption{Fully corrected $\Delta R$ distributions, measured with the $k_{\rm{T}}$, C/A and C/A with Soft Drop grooming algorithms, in pp collisions at $\sqrt{s} = 7$\,TeV for jets with $R = 0.4$ in 
the jet $p_{\rm T,\rm jet}^{\rm ch}$ interval of $40$--$60$\,GeV$/c$, are shown. The systematic uncertainties are given by the grey boxes. The results are compared with PYTHIA6 Perugia 2011 and PYTHIA8 Monash. The uncertainties presented for the PYTHIA distributions are purely statistical.}
\label{fig:Resultspp_DeltaR}
\end{figure}

\begin{figure}[h!]
\centering
\begin{minipage}[b]{0.49\textwidth}
\centering
\includegraphics[width=\textwidth]{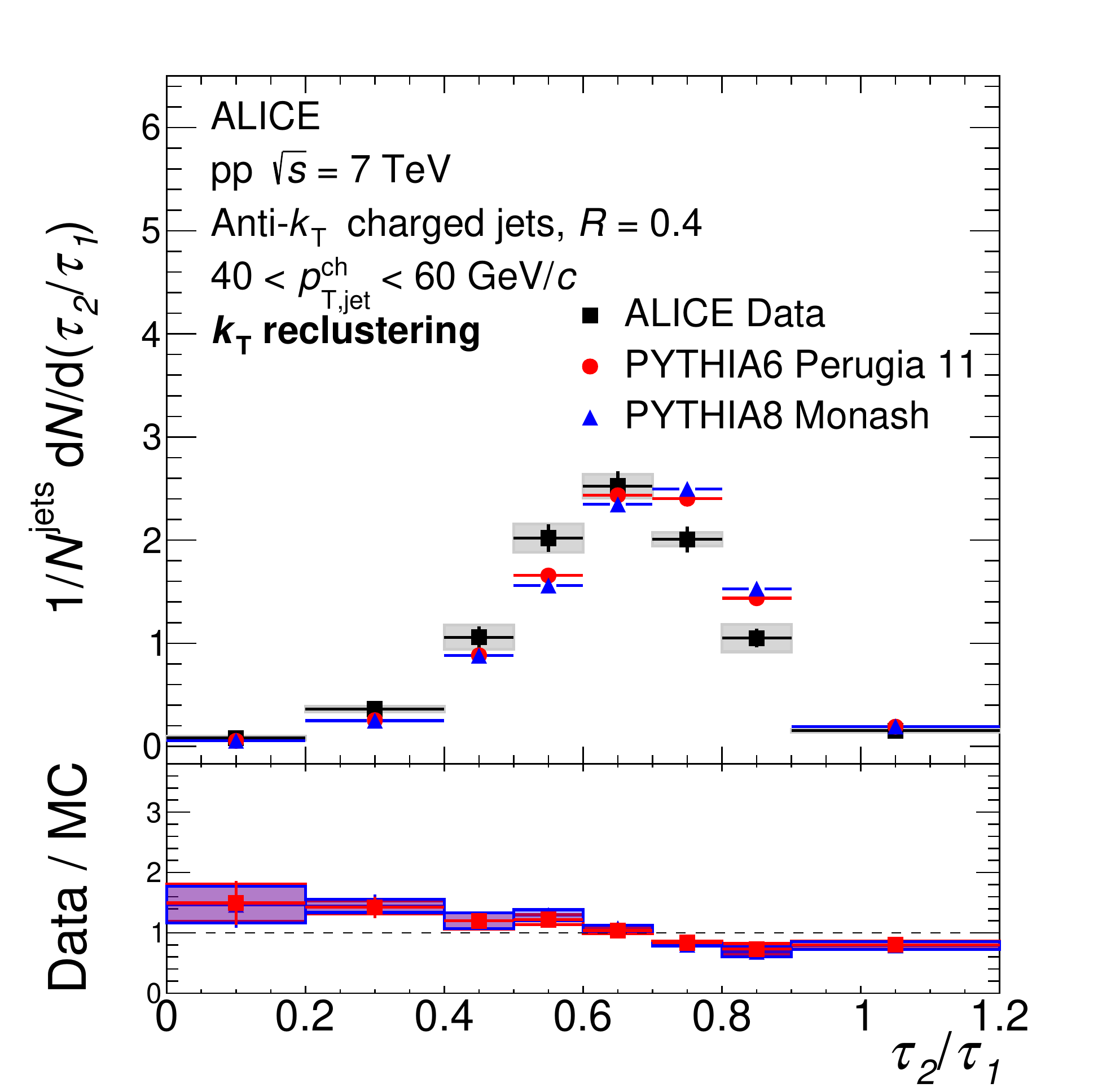}

\end{minipage}
\centering
\begin{minipage}[b]{0.49\textwidth}
\centering
\includegraphics[width=\textwidth]{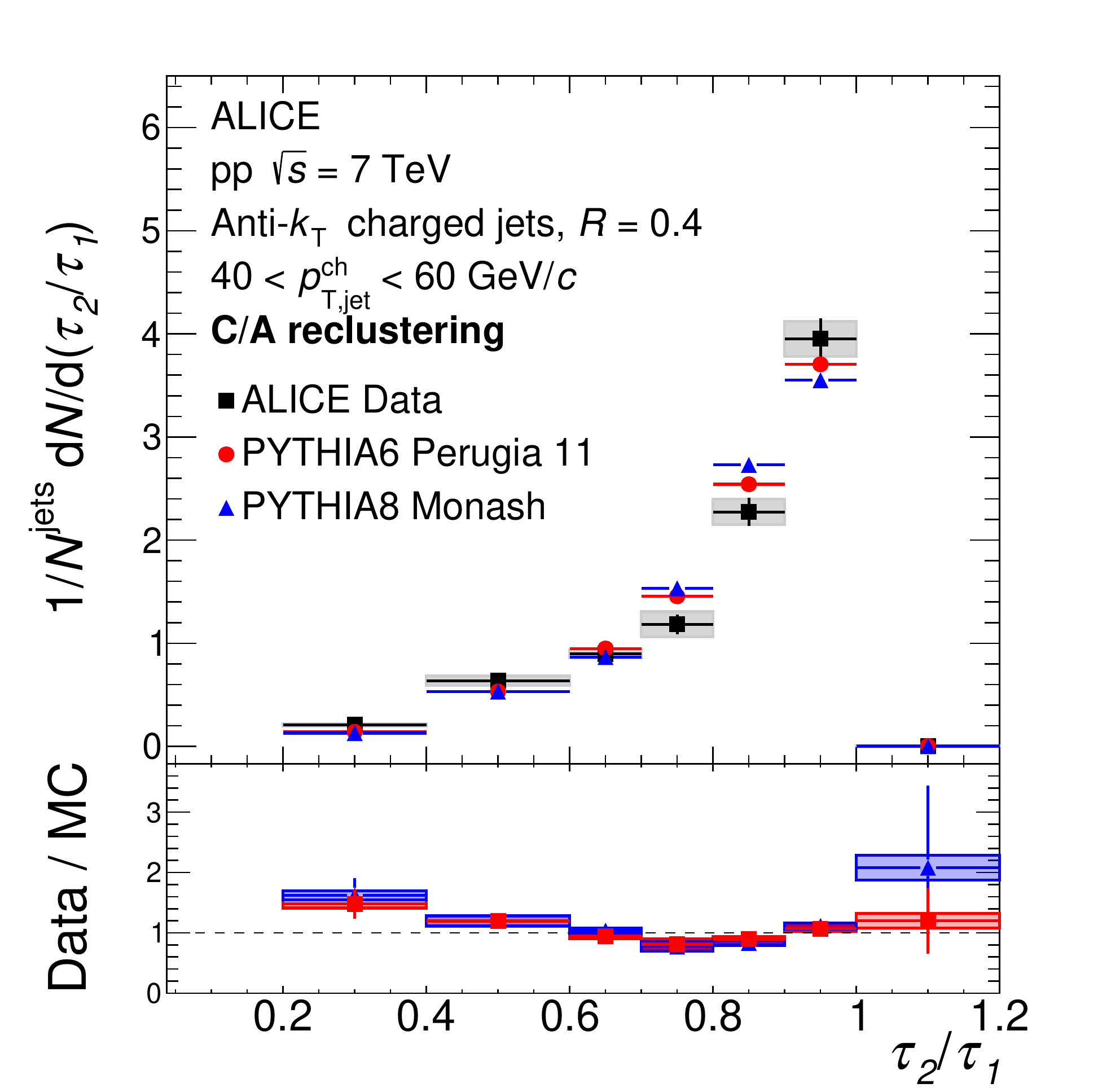}

\end{minipage}
\centering
\begin{minipage}[b]{0.49\textwidth}
\centering
\includegraphics[width=\textwidth]{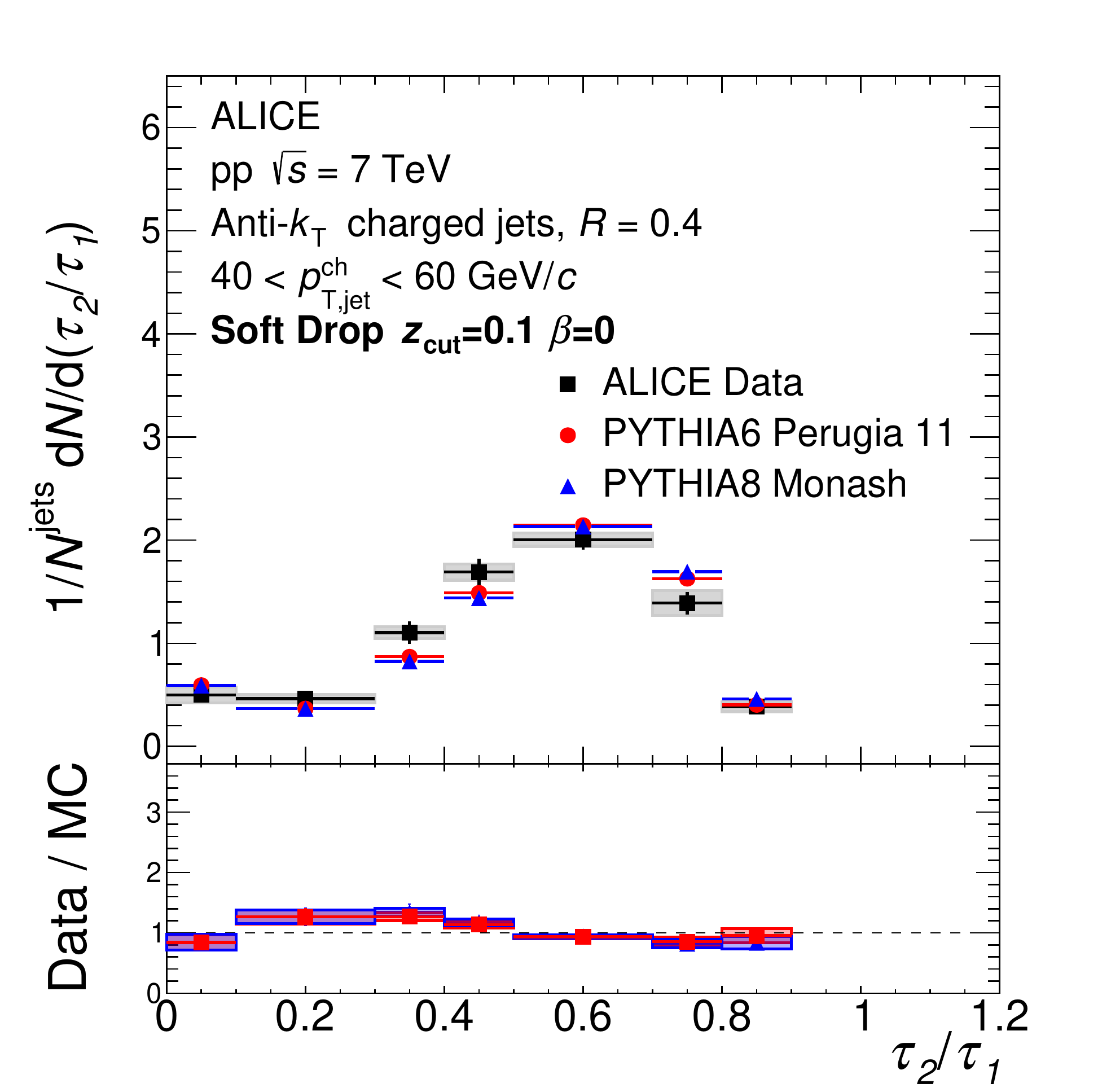}

\end{minipage}
\caption{Fully corrected $\tau_{2}/\tau_{1}$ distributions, measured with the $k_{\rm{T}}$, C/A and C/A with Soft Drop grooming algorithms, in pp collisions at $\sqrt{s} = 7$\,TeV for jets with $R = 0.4$ in 
the jet $p_{\rm T,\rm jet}^{\rm ch}$ interval of $40$--$60$\,GeV$/c$, are shown. The systematic uncertainties are given by the grey boxes. The results are compared with PYTHIA6 Perugia 2011 and PYTHIA8 Monash. The uncertainties presented for the PYTHIA distributions are purely statistical.}
\label{fig:Resultspp_Tau2to1}
\end{figure}

\noindent The fully corrected $\Delta R$ and $\tau_{2}/\tau_{1}$ distributions, measured in pp collisions at $\sqrt{s}=7$\,TeV in the $p_{\rm T,\rm jet}^{\rm ch}$ interval $40$--$60$\,GeV/$c$, are presented in Figs.~\ref{fig:Resultspp_DeltaR} and~\ref{fig:Resultspp_Tau2to1}, respectively. The C/A algorithm appears highly sensitive to the underlying event, with the axis of the subleading prong being strongly correlated to large angle (large $\Delta R$) soft (high $\tau_{2}/\tau_{1}$) radiation in the jet cone. The addition of the Soft Drop groomer modifies the subjet axes dramatically, with the axes being qualitatively similar to the $k_{\rm T}$ case. The lower $\tau_{2}/\tau_{1}$ values obtained with these reclustering algorithms, compared to the C/A case, show that both axes are aligned with a more significant fraction of the transverse momentum inside the jet. The $\Delta R$ distributions for these two algorithms peak at small values, showing that the jets measured in this sample are strongly single-cored. This is responsible for the $\tau_{2}/\tau_{1}$ distributions peaking at intermediate values, as the two hard substructures found in the jet are often not well separated and defined.

The results are compared with PYTHIA6 Perugia 2011 and PYTHIA8 Monash distributions obtained at the same collision energy. The opening angle between the subjet axes ($\Delta R$) is well described by the MC for all three reclustering algorithms. The fragmentation of particles around these axes ($\tau_{2}/\tau_{1}$) is less well described by the MC models. The PYTHIA distributions appear shifted towards less 2-prong jets compared to the data. Measurements of $\tau_{2}/\tau_{1}$ for top-tagged jets, made by the CMS collaboration~\cite{Sirunyan:2018asm}, show the same relation to MC models in a complementary jet momentum range. The results observed in pp collisions are used to validate the MC for comparisons with Pb--Pb. However, there are two caveats to consider during this validation. The first is the difference in $\sqrt{s}$ between the two collision systems. Due to the slow change in the underlying event in pp collisions, this is expected to be negligible. The second caveat is that measurements in pp collisions were performed on an inclusive jet sample, whilst those in Pb--Pb collisions were performed on jets recoiling from high-$p_{\rm T}$ trigger hadrons. These methods are incorporated into the respective MC calculations for each centre-of-mass energy. However, the recoiling sample of jets are expected to have no fragmentation biases, which would be the main concern when comparing the PYTHIA performances, on the two different samples, for the observable measured.

\begin{figure}[h!]
\centering
\begin{minipage}[b]{0.49\textwidth}
\centering
\includegraphics[width=\textwidth]{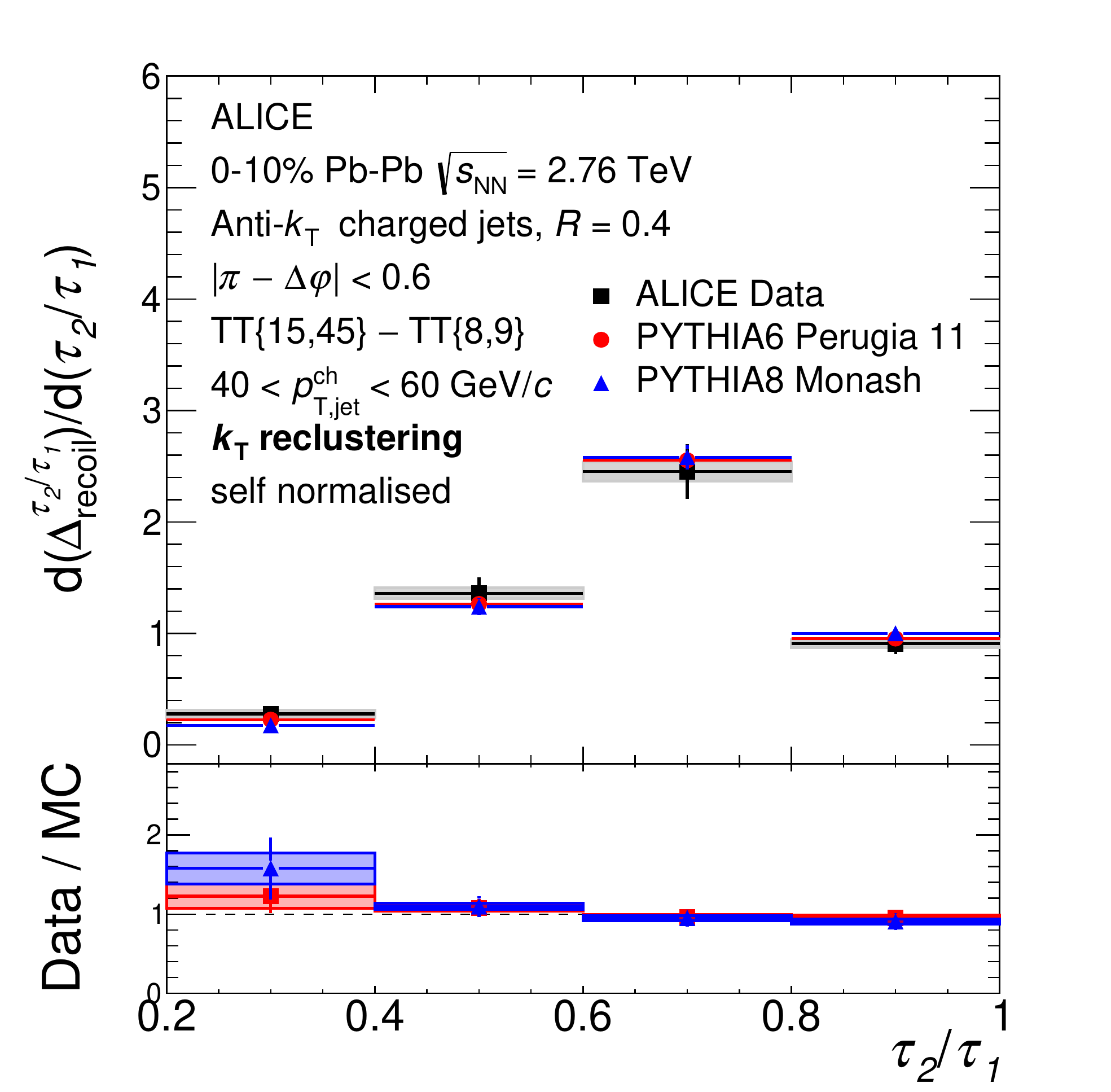}

\end{minipage}
\centering
\begin{minipage}[b]{0.49\textwidth}
\centering
\includegraphics[width=\textwidth]{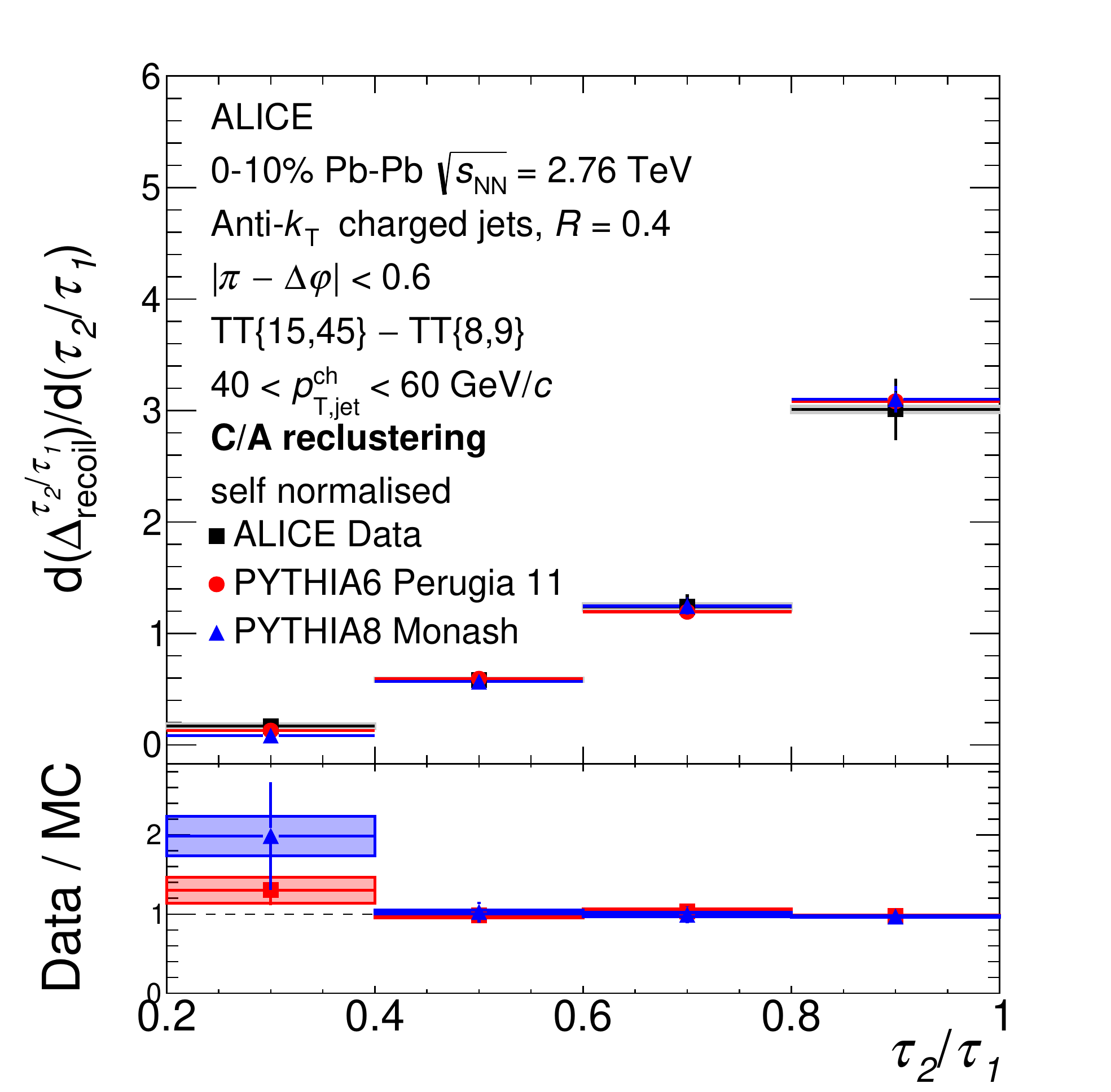}

\end{minipage}
\centering
\begin{minipage}[b]{0.49\textwidth}
\centering
\includegraphics[width=\textwidth]{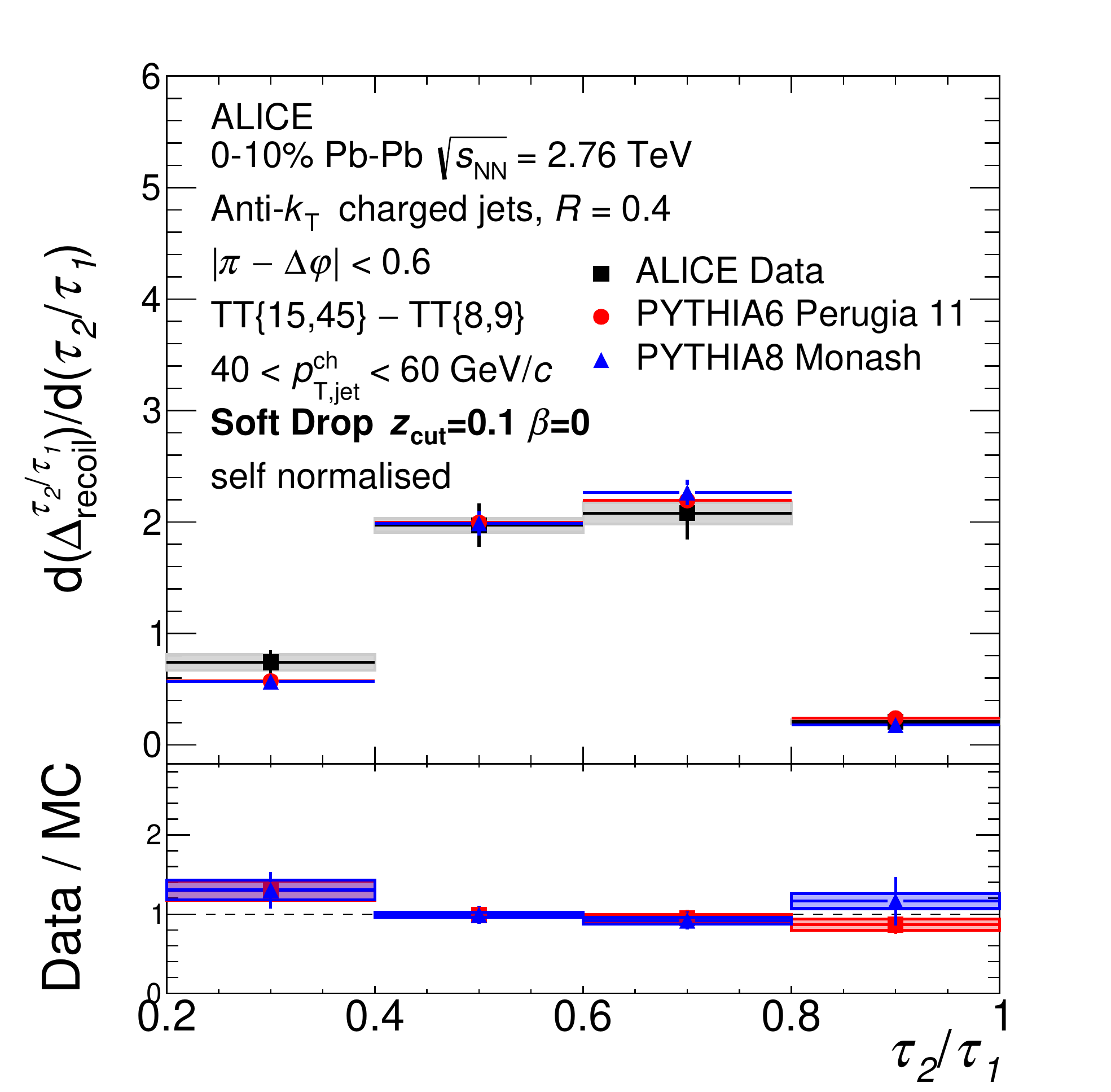}

\end{minipage}
\centering
\caption{Fully corrected $\tau_{2}/\tau_{1}$ distributions, measured with the $k_{\rm{T}}$, C/A and C/A with Soft Drop grooming algorithms, in Pb--Pb collisions at $\sqrt{s_{\rm NN}} = 2.76$\,TeV for jets with $R = 0.4$ in 
the jet $p_{\rm T,\rm jet}^{\rm ch}$ interval of $40$--$60$\,GeV$/c$, are shown. The systematic uncertainties are given by the grey boxes. The results are self normalised and compared with PYTHIA6 Perugia 2011 and PYTHIA8 Monash. The uncertainties presented for the PYTHIA distributions are purely statistical.}
\label{fig:ResultsPbPb_Tau2to1}
\end{figure}

\noindent Figure~\ref{fig:ResultsPbPb_Tau2to1} presents the fully corrected $\tau_{2}/\tau_{1}$ distributions measured in
Pb--Pb collisions at $\sqrt{s_{\rm NN}}=2.76$\,TeV in the $p_{\rm T,\rm jet}^{\rm ch}$ interval $40$--$60$\,GeV/$c$.  The results are compared with vacuum PYTHIA6 Perugia 2011 and PYTHIA8 Monash distributions obtained at the same collision energy. The ratio plots in the lower panels indicate an agreement of about $20\%$ between the data and MC. 
Since in pp collisions the MC seems to produce fewer jets with a distinct two-pronged substructure than data, the better agreement in Pb--Pb collisions alludes to the possibility that fewer two-prong jets are measured in heavy-ion collisions relative to pp, at the same reconstructed energy. This conclusion is most relevant for the $k_{\rm T}$ and Soft Drop algorithms, where the axes are aligned to hard substructures in the jet.

The measurement of the $\Delta R$ shape is not presented for Pb--Pb collisions. This is due to the non-diagonal nature of the $\Delta R$ response which arises from the presence of combinatorial subleading prongs, to which the $\Delta R$ shape is very sensitive. These prongs persist in a significant fraction of jets after the background subtraction procedure and contribute strongly in the region of phase space where the area is maximal, $\DeltaR \sim R$. This off-diagonality of the response renders the unfolding unstable and a fully corrected measurement cannot be presented. This response to the underlying event can be understood through the nature of the substructure observable in question, which can be broadly placed in one of two distinct classes. The first of these are observables which isolate a particular set of constituents or structures in the jet, such as a subjet, and then measure the properties of this isolated structure. Such observables are prone to non-continuous deformations by the underlying event, which in some cases are not possible to correct for through unfolding. $\Delta R$ is one such observable as it is a measure of the geometrical placement of subjets within the jet.

The second class of observables are those which do not isolate any part of the jet, but instead use all the tracks in the jet to statistically calculate a value, based on a given definition. The deformation of these observables by the background has a continuous characteristic and as such can be handled by the unfolding procedure.  $\tau_{2}/\tau_{1}$ is one such observable and is therefore not very sensitive to these combinatorial subleading prongs. In fact, a significant redistribution in the jet momentum is required to significantly alter the value of $\tau_{2}/\tau_{1}$. As the majority of the measured jet sample have a single-cored substructure, any potential displacement of the subleading axis, by a soft combinatorial axis, has minimal effect on the observable. In the $\tau_{2}$ case, the leading axis remains in the jet core, where the majority of the momentum is situated, and thus does not significantly alter the measured value. The $\tau_{1}$ value is also not significantly impacted as the axis remains close to the core and the contribution from the soft background to the calculation is small. Only the addition of a hard second prong has the potential to significantly alter the observable, by increasing the value of the $\tau_{1}$ variable. In this way, the $\tau_{2}/\tau_{1}$ observable is resilient to soft combinatorial prongs and can therefore be unfolded due to he diagonal nature of the response matrix.

 \section{Conclusions}
\label{sect:conclusions}

The first measurements of $\tau_{2}/\tau_{1}$ in heavy-ion collisions have been presented, as a means to explore a possible change in the degree to which the internal structure of jets are composed of two distinct substructures. This two-prongness of jets might be sensitive to coherence effects in the QGP, where jets with distinct substructures that are resolved by the medium, are expected to lose more energy compared to jets where the energy flow is concentrated in a single core. The measurements are made relative to a variety of axes choices, selected through the use of different reclustering metrics and grooming procedures, which are in turn potentially sensitive to different aspects of in-medium jet modification. In order to extend this substructure measurement to low jet transverse momentum and large jet resolution, where the impact of the underlying heavy-ion background on the yield of jets is large, a semi-inclusive hadron-jet coincidence technique was extended for the first time to a substructure observable, allowing for a fully corrected measurement in this unexplored regime. This sets the foundation for further such measurements in the future. Measurements reported in pp collisions provide both information on the alignment of QCD jet radiation relative to the different axes choices, as well as validating MC calculations for comparison with the Pb--Pb measurements. The aperture angle between the subjet axes, $\Delta \it{R}$, is also presented in pp collisions, providing complementary information on the spacial distinctness of the subjets returned by the different axes choices. This measurement also serves as a baseline for future measurements in heavy-ion collisions, which can be used to directly probe the coherence angle in the QGP.

In pp collisions, the MC calculations underestimate the two-prongness of jets (MC distributions of $\tau_{2}/\tau_{1}$ are shifted to larger values) whilst the aperture angle between the subjet axes remains well described. 
In Pb--Pb collisions, the measured two-prongness of jets is found to not be significantly modified relative to the MC reference, for a variety of different reclustering algorithms. These two findings hint at a reduction in the two-prongness of jets in heavy-ion collisions relative to pp collisions in the same measured jet momentum interval. 

Correlation studies performed by the CMS collaboration at particle level~\cite{Sirunyan:2018asm} indicate that the $\tau_{2}/\tau_{1}$ observable is weakly correlated with the majority of other substructure observables, measured at particle level, so far by the ALICE collaboration. These include the jet width, the jet momentum dispersion and the (Soft Drop) groomed jet radius. In this way, the presented $\tau_{2}/\tau_{1}$ measurements can provide extra constraints for jet quenching calculations and models. A relatively stronger correlation is observed between the $\tau_{2}/\tau_{1}$ observable and both the groomed momentum imbalance and the number of subjet prongs passing the Soft Drop condition. Detector-level measurements by the ALICE collaboration indicate a modification of the groomed momentum imbalance for wide-angle splittings~\cite{Acharya:2019djg}. Future fully corrected measurements of these observables can explore the three-dimensional correlations between the jet momentum, the groomed momentum imbalance and the groomed radius. In the same way, the correlation between the jet momentum, two-prongness of jets measured via $\tau_{2}/\tau_{1}$ and the apeture angle between the axes, can be used to understand the interplay between colour coherence and energy loss.

%%%%%%%%%%%%%%%%%%%%%%%%%%%%%%%%
% end main text 
%%%%%%%%%%%%%%%%%%%%%%%%%%%%%%%%

%%%%% acknowledgements - handled by EB chairs 
\newenvironment{acknowledgement}{\relax}{\relax}
\begin{acknowledgement}
\section*{Acknowledgements}
% add specific acknowledgements here 
% ...but please don't remove the line below: funding agencies
% will be acknowledged with a custom tex file handled by EB chairs after Collab Round 2
% Version: 2021-04-23

The ALICE Collaboration would like to thank all its engineers and technicians for their invaluable contributions to the construction of the experiment and the CERN accelerator teams for the outstanding performance of the LHC complex.
The ALICE Collaboration gratefully acknowledges the resources and support provided by all Grid centres and the Worldwide LHC Computing Grid (WLCG) collaboration.
The ALICE Collaboration acknowledges the following funding agencies for their support in building and running the ALICE detector:
A. I. Alikhanyan National Science Laboratory (Yerevan Physics Institute) Foundation (ANSL), State Committee of Science and World Federation of Scientists (WFS), Armenia;
Austrian Academy of Sciences, Austrian Science Fund (FWF): [M 2467-N36] and Nationalstiftung f\"{u}r Forschung, Technologie und Entwicklung, Austria;
Ministry of Communications and High Technologies, National Nuclear Research Center, Azerbaijan;
Conselho Nacional de Desenvolvimento Cient\'{\i}fico e Tecnol\'{o}gico (CNPq), Financiadora de Estudos e Projetos (Finep), Funda\c{c}\~{a}o de Amparo \`{a} Pesquisa do Estado de S\~{a}o Paulo (FAPESP) and Universidade Federal do Rio Grande do Sul (UFRGS), Brazil;
Ministry of Education of China (MOEC) , Ministry of Science \& Technology of China (MSTC) and National Natural Science Foundation of China (NSFC), China;
Ministry of Science and Education and Croatian Science Foundation, Croatia;
Centro de Aplicaciones Tecnol\'{o}gicas y Desarrollo Nuclear (CEADEN), Cubaenerg\'{\i}a, Cuba;
Ministry of Education, Youth and Sports of the Czech Republic, Czech Republic;
The Danish Council for Independent Research | Natural Sciences, the VILLUM FONDEN and Danish National Research Foundation (DNRF), Denmark;
Helsinki Institute of Physics (HIP), Finland;
Commissariat \`{a} l'Energie Atomique (CEA) and Institut National de Physique Nucl\'{e}aire et de Physique des Particules (IN2P3) and Centre National de la Recherche Scientifique (CNRS), France;
Bundesministerium f\"{u}r Bildung und Forschung (BMBF) and GSI Helmholtzzentrum f\"{u}r Schwerionenforschung GmbH, Germany;
General Secretariat for Research and Technology, Ministry of Education, Research and Religions, Greece;
National Research, Development and Innovation Office, Hungary;
Department of Atomic Energy Government of India (DAE), Department of Science and Technology, Government of India (DST), University Grants Commission, Government of India (UGC) and Council of Scientific and Industrial Research (CSIR), India;
Indonesian Institute of Science, Indonesia;
Istituto Nazionale di Fisica Nucleare (INFN), Italy;
Institute for Innovative Science and Technology , Nagasaki Institute of Applied Science (IIST), Japanese Ministry of Education, Culture, Sports, Science and Technology (MEXT) and Japan Society for the Promotion of Science (JSPS) KAKENHI, Japan;
Consejo Nacional de Ciencia (CONACYT) y Tecnolog\'{i}a, through Fondo de Cooperaci\'{o}n Internacional en Ciencia y Tecnolog\'{i}a (FONCICYT) and Direcci\'{o}n General de Asuntos del Personal Academico (DGAPA), Mexico;
Nederlandse Organisatie voor Wetenschappelijk Onderzoek (NWO), Netherlands;
The Research Council of Norway, Norway;
Commission on Science and Technology for Sustainable Development in the South (COMSATS), Pakistan;
Pontificia Universidad Cat\'{o}lica del Per\'{u}, Peru;
Ministry of Education and Science, National Science Centre and WUT ID-UB, Poland;
Korea Institute of Science and Technology Information and National Research Foundation of Korea (NRF), Republic of Korea;
Ministry of Education and Scientific Research, Institute of Atomic Physics and Ministry of Research and Innovation and Institute of Atomic Physics, Romania;
Joint Institute for Nuclear Research (JINR), Ministry of Education and Science of the Russian Federation, National Research Centre Kurchatov Institute, Russian Science Foundation and Russian Foundation for Basic Research, Russia;
Ministry of Education, Science, Research and Sport of the Slovak Republic, Slovakia;
National Research Foundation of South Africa, South Africa;
Swedish Research Council (VR) and Knut \& Alice Wallenberg Foundation (KAW), Sweden;
European Organization for Nuclear Research, Switzerland;
Suranaree University of Technology (SUT), National Science and Technology Development Agency (NSDTA) and Office of the Higher Education Commission under NRU project of Thailand, Thailand;
Turkish Energy, Nuclear and Mineral Research Agency (TENMAK), Turkey;
National Academy of  Sciences of Ukraine, Ukraine;
Science and Technology Facilities Council (STFC), United Kingdom;
National Science Foundation of the United States of America (NSF) and United States Department of Energy, Office of Nuclear Physics (DOE NP), United States of America.
\end{acknowledgement}

%%%%%%%% Bibliography 
\bibliographystyle{utphys}   % Remember we use title in the biblio
\bibliography{references.bib}

\providecommand{\href}[2]{#2}\begingroup\raggedright\begin{thebibliography}{10}

\bibitem{Muller:2006ee}
B.~Muller and J.~L. Nagle, ``{Results from the relativistic heavy ion
  collider}'',
  \href{http://dx.doi.org/10.1146/annurev.nucl.56.080805.140556}{{\em Ann. Rev.
  Nucl. Part. Sci.} {\bfseries 56} (2006) 93--135},
  \href{http://arxiv.org/abs/nucl-th/0602029}{{\ttfamily
  arXiv:nucl-th/0602029}}.

\bibitem{Roland:2014jsa}
G.~Roland, K.~Safarik, and P.~Steinberg, ``{Heavy-ion collisions at the LHC}'',
  \href{http://dx.doi.org/10.1016/j.ppnp.2014.05.001}{{\em Prog. Part. Nucl.
  Phys.} {\bfseries 77} (2014) 70--127}.

\bibitem{Dasgupta:2016bnd}
M.~Dasgupta, F.~A. Dreyer, G.~P. Salam, and G.~Soyez, ``{Inclusive jet spectrum
  for small-radius jets}'',
  \href{http://dx.doi.org/10.1007/JHEP06(2016)057}{{\em JHEP} {\bfseries 06}
  (2016) 057}, \href{http://arxiv.org/abs/1602.01110}{{\ttfamily
  arXiv:1602.01110 [hep-ph]}}.

\bibitem{Marzani:2019hun}
S.~Marzani, G.~Soyez, and M.~Spannowsky,
  \href{http://dx.doi.org/10.1007/978-3-030-15709-8}{{\em {Looking inside jets:
  an introduction to jet substructure and boosted-object phenomenology}}},
  vol.~958.
\newblock Springer, 2019.
\newblock \href{http://arxiv.org/abs/1901.10342}{{\ttfamily arXiv:1901.10342
  [hep-ph]}}.

\bibitem{Sirunyan:2018asm}
{\bfseries CMS} Collaboration, A.~M. Sirunyan {\em et~al.}, ``{Measurement of
  jet substructure observables in $\mathrm{t\overline{t}}$ events from
  proton-proton collisions at $\sqrt{s}=$ 13TeV}'',
  \href{http://dx.doi.org/10.1103/PhysRevD.98.092014}{{\em Phys. Rev. D}
  {\bfseries 98} no.~9, (2018) 092014},
  \href{http://arxiv.org/abs/1808.07340}{{\ttfamily arXiv:1808.07340
  [hep-ex]}}.

\bibitem{Larkoski:2017jix}
A.~J. Larkoski, I.~Moult, and B.~Nachman, ``{Jet Substructure at the Large
  Hadron Collider: A Review of Recent Advances in Theory and Machine
  Learning}'', \href{http://dx.doi.org/10.1016/j.physrep.2019.11.001}{{\em
  Phys. Rept.} {\bfseries 841} (2020) 1--63},
  \href{http://arxiv.org/abs/1709.04464}{{\ttfamily arXiv:1709.04464
  [hep-ph]}}.

\bibitem{Majumder:2010qh}
A.~Majumder and M.~Van~Leeuwen, ``{The Theory and Phenomenology of Perturbative
  QCD Based Jet Quenching}'',
  \href{http://dx.doi.org/10.1016/j.ppnp.2010.09.001}{{\em
  Prog.Part.Nucl.Phys.} {\bfseries A66} (2011) 41--92},
\href{http://arxiv.org/abs/1002.2206}{{\ttfamily arXiv:1002.2206 [hep-ph]}}.
%%CITATION = ARXIV:1002.2206;%%.

\bibitem{Adam:2015doa}
{\bfseries ALICE} Collaboration, J.~Adam {\em et~al.}, ``{Measurement of jet
  quenching with semi-inclusive hadron-jet distributions in central Pb--Pb
  collisions at $\sqrt{s_{\rm{NN}}} = 2.76$ TeV}'',
  \href{http://dx.doi.org/10.1007/JHEP09(2015)170}{{\em JHEP} {\bfseries 09}
  (2015) 170},
\href{http://arxiv.org/abs/1506.03984}{{\ttfamily arXiv:1506.03984 [nucl-ex]}}.
%%CITATION = ARXIV:1506.03984;%%.

\bibitem{Adamczyk:2017yhe}
{\bfseries STAR} Collaboration, L.~Adamczyk {\em et~al.}, ``{Measurements of
  jet quenching with semi-inclusive hadron+jet distributions in Au+Au
  collisions at $\sqrt{s_{NN}}$ = 200 GeV}'',
  \href{http://dx.doi.org/10.1103/PhysRevC.96.024905}{{\em Phys. Rev. C}
  {\bfseries 96} no.~2, (2017) 024905},
  \href{http://arxiv.org/abs/1702.01108}{{\ttfamily arXiv:1702.01108
  [nucl-ex]}}.

\bibitem{Sirunyan:2017jic}
{\bfseries CMS} Collaboration, A.~M. Sirunyan {\em et~al.}, ``{Study of Jet
  Quenching with $Z+\text{jet}$ Correlations in Pb-Pb and $pp$ Collisions at
  ${\sqrt{s}}_{NN}=5.02\text{ }\text{ }\mathrm{TeV}$}'',
  \href{http://dx.doi.org/10.1103/PhysRevLett.119.082301}{{\em Phys. Rev.
  Lett.} {\bfseries 119} no.~8, (2017) 082301},
  \href{http://arxiv.org/abs/1702.01060}{{\ttfamily arXiv:1702.01060
  [nucl-ex]}}.

\bibitem{Aaboud:2018anc}
{\bfseries ATLAS} Collaboration, M.~Aaboud {\em et~al.}, ``{Measurement of
  photon\textendash{}jet transverse momentum correlations in 5.02 TeV Pb + Pb
  and $pp$ collisions with ATLAS}'',
  \href{http://dx.doi.org/10.1016/j.physletb.2018.12.023}{{\em Phys. Lett. B}
  {\bfseries 789} (2019) 167--190},
  \href{http://arxiv.org/abs/1809.07280}{{\ttfamily arXiv:1809.07280
  [nucl-ex]}}.

\bibitem{Burke:2013yra}
K.~M. Burke, A.~Buzzatti, N.~Chang, C.~Gale, M.~Gyulassy, {\em et~al.},
  ``{Extracting jet transport coefficient from jet quenching at RHIC and
  LHC}'', \href{http://dx.doi.org/10.1103/PhysRevC.90.014909}{{\em Phys.Rev.}
  {\bfseries C90} (2014) 014909},
\href{http://arxiv.org/abs/1312.5003}{{\ttfamily arXiv:1312.5003 [nucl-th]}}.
%%CITATION = ARXIV:1312.5003;%%.

\bibitem{Acharya:2018uvf}
{\bfseries ALICE} Collaboration, S.~Acharya {\em et~al.}, ``{Medium
  modification of the shape of small-radius jets in central Pb-Pb collisions at
  $\sqrt{s_{\mathrm {NN}}} = 2.76\,\rm{TeV}$}'',
  \href{http://dx.doi.org/10.1007/JHEP10(2018)139}{{\em JHEP} {\bfseries 10}
  (2018) 139},
\href{http://arxiv.org/abs/1807.06854}{{\ttfamily arXiv:1807.06854 [nucl-ex]}}.
%%CITATION = ARXIV:1807.06854;%%.

\bibitem{Chatrchyan:2013kwa}
{\bfseries CMS} Collaboration, S.~Chatrchyan {\em et~al.}, ``{Modification of
  jet shapes in PbPb collisions at $\sqrt {s_{\rm NN}} = 2.76$ TeV}'',
  \href{http://dx.doi.org/10.1016/j.physletb.2014.01.042}{{\em Phys.Lett.}
  {\bfseries B730} (2014) 243--263},
\href{http://arxiv.org/abs/1310.0878}{{\ttfamily arXiv:1310.0878 [nucl-ex]}}.
%%CITATION = ARXIV:1310.0878;%%.

\bibitem{Sirunyan:2018qec}
{\bfseries CMS} Collaboration, A.~M. Sirunyan {\em et~al.}, ``{Observation of
  Medium-Induced Modifications of Jet Fragmentation in Pb-Pb Collisions at
  $\sqrt{s_{NN}}=$ 5.02 TeV Using Isolated Photon-Tagged Jets}'',
  \href{http://dx.doi.org/10.1103/PhysRevLett.121.242301}{{\em Phys. Rev.
  Lett.} {\bfseries 121} no.~24, (2018) 242301},
  \href{http://arxiv.org/abs/1801.04895}{{\ttfamily arXiv:1801.04895
  [hep-ex]}}.

\bibitem{Aaboud:2017bzv}
{\bfseries ATLAS} Collaboration, M.~Aaboud {\em et~al.}, ``{Measurement of jet
  fragmentation in Pb+Pb and $pp$ collisions at $\sqrt{{s_\mathrm{NN}}} = 2.76$
  TeV with the ATLAS detector at the LHC}'',
  \href{http://dx.doi.org/10.1140/epjc/s10052-017-4915-5}{{\em Eur. Phys. J.}
  {\bfseries C77} (2017) 379},
\href{http://arxiv.org/abs/1702.00674}{{\ttfamily arXiv:1702.00674 [hep-ex]}}.
%%CITATION = ARXIV:1702.00674;%%.

\bibitem{Acharya:2017goa}
{\bfseries ALICE} Collaboration, S.~Acharya {\em et~al.}, ``{First measurement
  of jet mass in Pb--Pb and p--Pb collisions at the LHC}'',
  \href{http://dx.doi.org/10.1016/j.physletb.2017.11.044}{{\em Phys. Lett.}
  {\bfseries B776} (2018) 249--264},
\href{http://arxiv.org/abs/1702.00804}{{\ttfamily arXiv:1702.00804 [nucl-ex]}}.
%%CITATION = ARXIV:1702.00804;%%.

\bibitem{Sirunyan:2018gct}
{\bfseries CMS} Collaboration, A.~M. Sirunyan {\em et~al.}, ``{Measurement of
  the groomed jet mass in PbPb and pp collisions at $
  \sqrt{s_{\mathrm{NN}}}=5.02 $ TeV}'',
  \href{http://dx.doi.org/10.1007/JHEP10(2018)161}{{\em JHEP} {\bfseries 10}
  (2018) 161}, \href{http://arxiv.org/abs/1805.05145}{{\ttfamily
  arXiv:1805.05145 [hep-ex]}}.

\bibitem{Sirunyan:2017bsd}
{\bfseries CMS} Collaboration, A.~M. Sirunyan {\em et~al.}, ``{Measurement of
  the Splitting Function in $pp$ and Pb-Pb Collisions at
  $\sqrt{s_{_{\mathrm{NN}}}} =$ 5.02 TeV}'',
  \href{http://dx.doi.org/10.1103/PhysRevLett.120.142302}{{\em Phys. Rev.
  Lett.} {\bfseries 120} no.~14, (2018) 142302},
  \href{http://arxiv.org/abs/1708.09429}{{\ttfamily arXiv:1708.09429
  [nucl-ex]}}.

\bibitem{Kauder:2017mhg}
{\bfseries STAR} Collaboration, K.~Kauder, ``{Measurement of the Shared
  Momentum Fraction $z_{\rm g}$ using Jet Reconstruction in p+p and Au+Au
  Collisions with STAR}'',
  \href{http://dx.doi.org/10.1016/j.nuclphysa.2017.07.004}{{\em Nucl. Phys.}
  {\bfseries A967} (2017) 516--519},
\href{http://arxiv.org/abs/1704.03046}{{\ttfamily arXiv:1704.03046 [nucl-ex]}}.
%%CITATION = ARXIV:1704.03046;%%.

\bibitem{Acharya:2019djg}
{\bfseries ALICE} Collaboration, S.~Acharya {\em et~al.}, ``{Exploration of jet
  substructure using iterative declustering in pp and Pb--Pb collisions at LHC
  energies}'', \href{http://dx.doi.org/10.1016/j.physletb.2020.135227}{{\em
  Phys. Lett. B} {\bfseries 802} (2020) 135227},
  \href{http://arxiv.org/abs/1905.02512}{{\ttfamily arXiv:1905.02512
  [nucl-ex]}}.

\bibitem{CasalderreySolana:2012ef}
J.~Casalderrey-Solana, Y.~Mehtar-Tani, C.~A. Salgado, and K.~Tywoniuk, ``{New
  picture of jet quenching dictated by color coherence}'',
  \href{http://dx.doi.org/10.1016/j.physletb.2013.07.046}{{\em Phys. Lett.}
  {\bfseries B725} (2013) 357--360},
\href{http://arxiv.org/abs/1210.7765}{{\ttfamily arXiv:1210.7765 [hep-ph]}}.
%%CITATION = ARXIV:1210.7765;%%.

\bibitem{D'Eramo:2012jh}
F.~D'Eramo, M.~Lekaveckas, H.~Liu, and K.~Rajagopal, ``{Momentum Broadening in
  Weakly Coupled Quark-Gluon Plasma (with a view to finding the quasiparticles
  within liquid quark-gluon plasma)}'',
  \href{http://dx.doi.org/10.1007/JHEP05(2013)031}{{\em JHEP} {\bfseries 05}
  (2013) 031},
\href{http://arxiv.org/abs/1211.1922}{{\ttfamily arXiv:1211.1922 [hep-ph]}}.
%%CITATION = ARXIV:1211.1922;%%.

\bibitem{Butterworth:2008iy}
J.~M. Butterworth, A.~R. Davison, M.~Rubin, and G.~P. Salam, ``{Jet
  substructure as a new Higgs search channel at the LHC}'',
  \href{http://dx.doi.org/10.1103/PhysRevLett.100.242001}{{\em Phys. Rev.
  Lett.} {\bfseries 100} (2008) 242001},
\href{http://arxiv.org/abs/0802.2470}{{\ttfamily arXiv:0802.2470 [hep-ph]}}.
%%CITATION = ARXIV:0802.2470;%%.

\bibitem{Larkoski:2014wba}
A.~J. Larkoski, S.~Marzani, G.~Soyez, and J.~Thaler, ``{Soft Drop}'',
  \href{http://dx.doi.org/10.1007/JHEP05(2014)146}{{\em JHEP} {\bfseries 05}
  (2014) 146},
\href{http://arxiv.org/abs/1402.2657}{{\ttfamily arXiv:1402.2657 [hep-ph]}}.
%%CITATION = ARXIV:1402.2657;%%.

\bibitem{Andrews:2018jcm}
H.~A. Andrews {\em et~al.}, ``{Novel tools and observables for jet physics in
  heavy-ion collisions}'',
  \href{http://dx.doi.org/10.1088/1361-6471/ab7cbc}{{\em J. Phys. G} {\bfseries
  47} no.~6, (2020) 065102}, \href{http://arxiv.org/abs/1808.03689}{{\ttfamily
  arXiv:1808.03689 [hep-ph]}}.

\bibitem{Sjostrand:2006za}
T.~Sj{\"o}strand, S.~Mrenna, and P.~Z. Skands, ``{PYTHIA 6.4 Physics and
  Manual}'', \href{http://dx.doi.org/10.1088/1126-6708/2006/05/026}{{\em JHEP}
  {\bfseries 05} (2006) 026},
\href{http://arxiv.org/abs/hep-ph/0603175}{{\ttfamily arXiv:hep-ph/0603175
  [hep-ph]}}.
%%CITATION = HEP-PH/0603175;%%.

\bibitem{FastJetAntikt}
M.~Cacciari, G.~P. Salam, and G.~Soyez, ``{The anti-$k_t$ jet clustering
  algorithm}'', \href{http://dx.doi.org/10.1088/1126-6708/2008/04/063}{{\em
  JHEP} {\bfseries 04} (2008) 063},
\href{http://arxiv.org/abs/0802.1189}{{\ttfamily arXiv:0802.1189 [hep-ph]}}.
%%CITATION = 0802.1189;%%.

\bibitem{Thaler:2010tr}
J.~Thaler and K.~Van~Tilburg, ``{Identifying Boosted Objects with
  N-subjettiness}'', \href{http://dx.doi.org/10.1007/JHEP03(2011)015}{{\em
  JHEP} {\bfseries 03} (2011) 015},
\href{http://arxiv.org/abs/1011.2268}{{\ttfamily arXiv:1011.2268 [hep-ph]}}.
%%CITATION = ARXIV:1011.2268;%%.

\bibitem{CMS-PAS-JME-13-007}
{\bfseries CMS} Collaboration, ``{Boosted Top Jet Tagging at CMS}'', tech.
  rep., CERN, Geneva, 2014.
\newblock \url{https://cds.cern.ch/record/1647419}.

\bibitem{Mehtar-Tani:2016aco}
Y.~Mehtar-Tani and K.~Tywoniuk, ``{Groomed jets in heavy-ion collisions:
  sensitivity to medium-induced bremsstrahlung}'',
  \href{http://dx.doi.org/10.1007/JHEP04(2017)125}{{\em JHEP} {\bfseries 04}
  (2017) 125},
\href{http://arxiv.org/abs/1610.08930}{{\ttfamily arXiv:1610.08930 [hep-ph]}}.
%%CITATION = ARXIV:1610.08930;%%.

\bibitem{Ellis:1993tq}
S.~D. Ellis and D.~E. Soper, ``{Successive combination jet algorithm for hadron
  collisions}'', \href{http://dx.doi.org/10.1103/PhysRevD.48.3160}{{\em Phys.
  Rev.} {\bfseries D48} (1993) 3160--3166},
\href{http://arxiv.org/abs/hep-ph/9305266}{{\ttfamily arXiv:hep-ph/9305266
  [hep-ph]}}.
%%CITATION = HEP-PH/9305266;%%.

\bibitem{Dokshitzer:1997in}
Y.~L. Dokshitzer, G.~D. Leder, S.~Moretti, and B.~R. Webber, ``{Better jet
  clustering algorithms}'',
  \href{http://dx.doi.org/10.1088/1126-6708/1997/08/001}{{\em JHEP} {\bfseries
  08} (1997) 001},
\href{http://arxiv.org/abs/hep-ph/9707323}{{\ttfamily arXiv:hep-ph/9707323
  [hep-ph]}}.
%%CITATION = HEP-PH/9707323;%%.

\bibitem{Dasgupta:2013ihk}
M.~Dasgupta, A.~Fregoso, S.~Marzani, and G.~P. Salam, ``{Towards an
  understanding of jet substructure}'',
  \href{http://dx.doi.org/10.1007/JHEP09(2013)029}{{\em JHEP} {\bfseries 09}
  (2013) 029}, \href{http://arxiv.org/abs/1307.0007}{{\ttfamily arXiv:1307.0007
  [hep-ph]}}.

\bibitem{Thaler:2011gf}
J.~Thaler and K.~Van~Tilburg, ``{Maximizing Boosted Top Identification by
  Minimizing N-subjettiness}'',
  \href{http://dx.doi.org/10.1007/JHEP02(2012)093}{{\em JHEP} {\bfseries 02}
  (2012) 093},
\href{http://arxiv.org/abs/1108.2701}{{\ttfamily arXiv:1108.2701 [hep-ph]}}.
%%CITATION = ARXIV:1108.2701;%%.

\bibitem{Aamodt:2008zz}
{\bfseries ALICE} Collaboration, K.~Aamodt {\em et~al.}, ``{The ALICE
  experiment at the CERN LHC}'',
\href{http://dx.doi.org/10.1088/1748-0221/3/08/S08002}{{\em JINST} {\bfseries
  3} (2008) S08002}.
%%CITATION = JINST,3,S08002;%%.

\bibitem{Abelev:2014ffa}
{\bfseries ALICE} Collaboration, B.~B. Abelev {\em et~al.}, ``{Performance of
  the ALICE Experiment at the CERN LHC}'',
  \href{http://dx.doi.org/10.1142/S0217751X14300440}{{\em Int. J. Mod. Phys.}
  {\bfseries A29} (2014) 1430044},
\href{http://arxiv.org/abs/1402.4476}{{\ttfamily arXiv:1402.4476 [nucl-ex]}}.
%%CITATION = ARXIV:1402.4476;%%.

\bibitem{ALICE:2014dla}
{\bfseries ALICE} Collaboration, B.~B. Abelev {\em et~al.}, ``{Charged jet
  cross sections and properties in proton-proton collisions at $\sqrt{s}=7$
  TeV}'', \href{http://dx.doi.org/10.1103/PhysRevD.91.112012}{{\em Phys. Rev.}
  {\bfseries D91} (2015) 112012},
\href{http://arxiv.org/abs/1411.4969}{{\ttfamily arXiv:1411.4969 [nucl-ex]}}.
%%CITATION = ARXIV:1411.4969;%%.

\bibitem{Abbas:2013taa}
{\bfseries ALICE} Collaboration, E.~Abbas {\em et~al.}, ``{Performance of the
  ALICE VZERO system}'',
\href{http://dx.doi.org/10.1088/1748-0221/8/10/P10016}{{\em JINST} {\bfseries
  8} (2013) P10016}.
%%CITATION = ARXIV:1306.3130;%%.

\bibitem{Aamodt:2010aa}
{\bfseries ALICE} Collaboration, K.~Aamodt {\em et~al.}, ``{Alignment of the
  ALICE Inner Tracking System with cosmic-ray tracks}'',
\href{http://dx.doi.org/10.1088/1748-0221/5/03/P03003}{{\em JINST} {\bfseries
  5} (2010) P03003}.
%%CITATION = ARXIV:1001.0502;%%.

\bibitem{Alme:2010ke}
J.~Alme {\em et~al.}, ``{The ALICE TPC, a large 3-dimensional tracking device
  with fast readout for ultra-high multiplicity events}'',
  \href{http://dx.doi.org/10.1016/j.nima.2010.04.042}{{\em Nucl. Instrum. Meth.
  A} {\bfseries 622} (2010) 316--367},
  \href{http://arxiv.org/abs/1001.1950}{{\ttfamily arXiv:1001.1950
  [physics.ins-det]}}.

\bibitem{Adam:2015ewa}
{\bfseries ALICE} Collaboration, J.~Adam {\em et~al.}, ``{Measurement of jet
  suppression in central Pb-Pb collisions at $\sqrt{s_{\rm NN}}$ = 2.76 TeV}'',
  \href{http://dx.doi.org/10.1016/j.physletb.2015.04.039}{{\em Phys. Lett. B}
  {\bfseries 746} (2015) 1--14},
  \href{http://arxiv.org/abs/1502.01689}{{\ttfamily arXiv:1502.01689
  [nucl-ex]}}.

\bibitem{Abelev:2012hxa}
{\bfseries ALICE} Collaboration, B.~Abelev {\em et~al.}, ``{Centrality
  Dependence of Charged Particle Production at Large Transverse Momentum in
  Pb--Pb Collisions at $\sqrt{s_{\rm{NN}}} = 2.76$ TeV}'',
  \href{http://dx.doi.org/10.1016/j.physletb.2013.01.051}{{\em Phys. Lett.}
  {\bfseries B720} (2013) 52--62},
\href{http://arxiv.org/abs/1208.2711}{{\ttfamily arXiv:1208.2711 [hep-ex]}}.
%%CITATION = ARXIV:1208.2711;%%.

\bibitem{Skands:2010ak}
P.~Z. Skands, ``{Tuning Monte Carlo Generators: The Perugia Tunes}'',
  \href{http://dx.doi.org/10.1103/PhysRevD.82.074018}{{\em Phys.Rev.}
  {\bfseries D82} (2010) 074018},
\href{http://arxiv.org/abs/1005.3457}{{\ttfamily arXiv:1005.3457 [hep-ph]}}.
%%CITATION = ARXIV:1005.3457;%%.

\bibitem{GEANT3}
{R. Brun, F. Bruyant, M. Maire, A.C. McPherson, and P. Zanarini}, ``{GEANT3
  User's Guide}'', {\em CERN Data Handling Division DD/EE/84-1} (1985) .

\bibitem{Cacciari:2011ma}
M.~Cacciari, G.~P. Salam, and G.~Soyez, ``{FastJet User Manual}'',
  \href{http://dx.doi.org/10.1140/epjc/s10052-012-1896-2}{{\em Eur.Phys.J.}
  {\bfseries C72} (2012) 1896},
\href{http://arxiv.org/abs/1111.6097}{{\ttfamily arXiv:1111.6097 [hep-ph]}}.
%%CITATION = ARXIV:1111.6097;%%.

\bibitem{FastJetArea}
M.~Cacciari, G.~P. Salam, and G.~Soyez, ``{The Catchment Area of Jets}'',
  \href{http://dx.doi.org/10.1088/1126-6708/2008/04/005}{{\em JHEP} {\bfseries
  04} (2008) 005},
\href{http://arxiv.org/abs/0802.1188}{{\ttfamily arXiv:0802.1188 [hep-ph]}}.
%%CITATION = 0802.1188;%%.

\bibitem{Jacobs:2010wq}
{\bfseries STAR} Collaboration, P.~M. Jacobs, ``{Background Fluctuations in
  Heavy Ion Jet Reconstruction}'',
  \href{http://dx.doi.org/10.1016/j.nuclphysa.2011.02.064}{{\em Nucl. Phys.}
  {\bfseries A855} (2011) 299},
\href{http://arxiv.org/abs/1012.2406}{{\ttfamily arXiv:1012.2406 [nucl-ex]}}.
%%CITATION = ARXIV:1012.2406;%%.

\bibitem{Berta:2014eza}
P.~Berta, M.~Spousta, D.~W. Miller, and R.~Leitner, ``{Particle-level pileup
  subtraction for jets and jet shapes}'',
  \href{http://dx.doi.org/10.1007/JHEP06(2014)092}{{\em JHEP} {\bfseries 06}
  (2014) 092},
\href{http://arxiv.org/abs/1403.3108}{{\ttfamily arXiv:1403.3108 [hep-ex]}}.
%%CITATION = ARXIV:1403.3108;%%.

\bibitem{Soyez:2012hv}
G.~Soyez, G.~P. Salam, J.~Kim, S.~Dutta, and M.~Cacciari, ``{Pileup subtraction
  for jet shapes}'',
  \href{http://dx.doi.org/10.1103/PhysRevLett.110.162001}{{\em Phys. Rev.
  Lett.} {\bfseries 110} (2013) 162001},
\href{http://arxiv.org/abs/1211.2811}{{\ttfamily arXiv:1211.2811 [hep-ph]}}.
%%CITATION = ARXIV:1211.2811;%%.

\bibitem{RooUnfold}
``Roounfold webpage.''
  \url{http://hepunx.rl.ac.uk/~adye/software/unfold/RooUnfold.html}.

\end{thebibliography}\endgroup

%%%%%%%%%%%%%%%%%%%%%%%%%%%%%%%%
% Appendices: yours (if any) + authorlist
%%%%%%%%%%%%%%%%%%%%%%%%%%%%%%%%
\newpage
\appendix

%
%\input{} % put your appendices here (if any)
%

%%%%% Authorlist - please do not touch: handled by EB chairs 
\section{The ALICE Collaboration}
\label{app:collab}
% ALICE Collaboration author list for 2021-04-23
%\documentclass[11pt, a4paper]{article} 
%\usepackage{geometry} 
%\begin{document} 
%\begin{flushleft} 
%\textbf{\Large ALICE Collaboration} 
%
%\bigskip 
\begingroup
\small
\begin{flushleft}
S.~Acharya$^{\rm 143}$, 
D.~Adamov\'{a}$^{\rm 98}$, 
A.~Adler$^{\rm 76}$, 
J.~Adolfsson$^{\rm 83}$, 
G.~Aglieri Rinella$^{\rm 35}$, 
M.~Agnello$^{\rm 31}$, 
N.~Agrawal$^{\rm 55}$, 
Z.~Ahammed$^{\rm 143}$, 
S.~Ahmad$^{\rm 16}$, 
S.U.~Ahn$^{\rm 78}$
% ...

\section*{Affiliation notes}

$^{\rm I}$ Deceased\\
$^{\rm II}$ Also at: Italian National Agency for New Technologies, Energy and Sustainable Economic Development (ENEA), Bologna, Italy\\
$^{\rm III}$ Also at: Dipartimento DET del Politecnico di Torino, Turin, Italy\\
$^{\rm IV}$ Also at: M.V. Lomonosov Moscow State University, D.V. Skobeltsyn Institute of Nuclear, Physics, Moscow, Russia\\
$^{\rm V}$ Also at: Institute of Theoretical Physics, University of Wroclaw, Poland\\

\section*{Collaboration Institutes}

$^{1}$ A.I. Alikhanyan National Science Laboratory (Yerevan Physics Institute) Foundation, Yerevan, Armenia\\
$^{2}$ AGH University of Science and Technology, Cracow, Poland\\
$^{3}$ Bogolyubov Institute for Theoretical Physics, National Academy of Sciences of Ukraine, Kiev, Ukraine\\
$^{4}$ Bose Institute, Department of Physics  and Centre for Astroparticle Physics and Space Science (CAPSS), Kolkata, India\\
$^{5}$ Budker Institute for Nuclear Physics, Novosibirsk, Russia\\
$^{6}$ California Polytechnic State University, San Luis Obispo, California, United States\\
$^{7}$ Central China Normal University, Wuhan, China\\
$^{8}$ Centro de Aplicaciones Tecnol\'{o}gicas y Desarrollo Nuclear (CEADEN), Havana, Cuba\\
$^{9}$ Centro de Investigaci\'{o}n y de Estudios Avanzados (CINVESTAV), Mexico City and M\'{e}rida, Mexico\\
$^{10}$ Chicago State University, Chicago, Illinois, United States\\
$^{11}$ China Institute of Atomic Energy, Beijing, China\\
$^{12}$ Chungbuk National University, Cheongju, Republic of Korea\\
$^{13}$ Comenius University Bratislava, Faculty of Mathematics, Physics and Informatics, Bratislava, Slovakia\\
$^{14}$ COMSATS University Islamabad, Islamabad, Pakistan\\
$^{15}$ Creighton University, Omaha, Nebraska, United States\\
$^{16}$ Department of Physics, Aligarh Muslim University, Aligarh, India\\
$^{17}$ Department of Physics, Pusan National University, Pusan, Republic of Korea\\
$^{18}$ Department of Physics, Sejong University, Seoul, Republic of Korea\\
$^{19}$ Department of Physics, University of California, Berkeley, California, United States\\
$^{20}$ Department of Physics, University of Oslo, Oslo, Norway\\
$^{21}$ Department of Physics and Technology, University of Bergen, Bergen, Norway\\
$^{22}$ Dipartimento di Fisica dell'Universit\`{a} 'La Sapienza' and Sezione INFN, Rome, Italy\\
$^{23}$ Dipartimento di Fisica dell'Universit\`{a} and Sezione INFN, Cagliari, Italy\\
$^{24}$ Dipartimento di Fisica dell'Universit\`{a} and Sezione INFN, Trieste, Italy\\
$^{25}$ Dipartimento di Fisica dell'Universit\`{a} and Sezione INFN, Turin, Italy\\
$^{26}$ Dipartimento di Fisica e Astronomia dell'Universit\`{a} and Sezione INFN, Bologna, Italy\\
$^{27}$ Dipartimento di Fisica e Astronomia dell'Universit\`{a} and Sezione INFN, Catania, Italy\\
$^{28}$ Dipartimento di Fisica e Astronomia dell'Universit\`{a} and Sezione INFN, Padova, Italy\\
$^{29}$ Dipartimento di Fisica e Nucleare e Teorica, Universit\`{a} di Pavia, Pavia, Italy\\
$^{30}$ Dipartimento di Fisica `E.R.~Caianiello' dell'Universit\`{a} and Gruppo Collegato INFN, Salerno, Italy\\
$^{31}$ Dipartimento DISAT del Politecnico and Sezione INFN, Turin, Italy\\
$^{32}$ Dipartimento di Scienze e Innovazione Tecnologica dell'Universit\`{a} del Piemonte Orientale and INFN Sezione di Torino, Alessandria, Italy\\
$^{33}$ Dipartimento di Scienze MIFT, Universit\`{a} di Messina, Messina, Italy\\
$^{34}$ Dipartimento Interateneo di Fisica `M.~Merlin' and Sezione INFN, Bari, Italy\\
$^{35}$ European Organization for Nuclear Research (CERN), Geneva, Switzerland\\
$^{36}$ Faculty of Electrical Engineering, Mechanical Engineering and Naval Architecture, University of Split, Split, Croatia\\
$^{37}$ Faculty of Engineering and Science, Western Norway University of Applied Sciences, Bergen, Norway\\
$^{38}$ Faculty of Nuclear Sciences and Physical Engineering, Czech Technical University in Prague, Prague, Czech Republic\\
$^{39}$ Faculty of Science, P.J.~\v{S}af\'{a}rik University, Ko\v{s}ice, Slovakia\\
$^{40}$ Frankfurt Institute for Advanced Studies, Johann Wolfgang Goethe-Universit\"{a}t Frankfurt, Frankfurt, Germany\\
$^{41}$ Fudan University, Shanghai, China\\
$^{42}$ Gangneung-Wonju National University, Gangneung, Republic of Korea\\
$^{43}$ Gauhati University, Department of Physics, Guwahati, India\\
$^{44}$ Helmholtz-Institut f\"{u}r Strahlen- und Kernphysik, Rheinische Friedrich-Wilhelms-Universit\"{a}t Bonn, Bonn, Germany\\
$^{45}$ Helsinki Institute of Physics (HIP), Helsinki, Finland\\
$^{46}$ High Energy Physics Group,  Universidad Aut\'{o}noma de Puebla, Puebla, Mexico\\
$^{47}$ Hiroshima University, Hiroshima, Japan\\
$^{48}$ Hochschule Worms, Zentrum  f\"{u}r Technologietransfer und Telekommunikation (ZTT), Worms, Germany\\
$^{49}$ Horia Hulubei National Institute of Physics and Nuclear Engineering, Bucharest, Romania\\
$^{50}$ Indian Institute of Technology Bombay (IIT), Mumbai, India\\
$^{51}$ Indian Institute of Technology Indore, Indore, India\\
$^{52}$ Indonesian Institute of Sciences, Jakarta, Indonesia\\
$^{53}$ INFN, Laboratori Nazionali di Frascati, Frascati, Italy\\
$^{54}$ INFN, Sezione di Bari, Bari, Italy\\
$^{55}$ INFN, Sezione di Bologna, Bologna, Italy\\
$^{56}$ INFN, Sezione di Cagliari, Cagliari, Italy\\
$^{57}$ INFN, Sezione di Catania, Catania, Italy\\
$^{58}$ INFN, Sezione di Padova, Padova, Italy\\
$^{59}$ INFN, Sezione di Pavia, Pavia, Italy\\
$^{60}$ INFN, Sezione di Roma, Rome, Italy\\
$^{61}$ INFN, Sezione di Torino, Turin, Italy\\
$^{62}$ INFN, Sezione di Trieste, Trieste, Italy\\
$^{63}$ Inha University, Incheon, Republic of Korea\\
$^{64}$ Institute for Gravitational and Subatomic Physics (GRASP), Utrecht University/Nikhef, Utrecht, Netherlands\\
$^{65}$ Institute for Nuclear Research, Academy of Sciences, Moscow, Russia\\
$^{66}$ Institute of Experimental Physics, Slovak Academy of Sciences, Ko\v{s}ice, Slovakia\\
$^{67}$ Institute of Physics, Homi Bhabha National Institute, Bhubaneswar, India\\
$^{68}$ Institute of Physics of the Czech Academy of Sciences, Prague, Czech Republic\\
$^{69}$ Institute of Space Science (ISS), Bucharest, Romania\\
$^{70}$ Institut f\"{u}r Kernphysik, Johann Wolfgang Goethe-Universit\"{a}t Frankfurt, Frankfurt, Germany\\
$^{71}$ Instituto de Ciencias Nucleares, Universidad Nacional Aut\'{o}noma de M\'{e}xico, Mexico City, Mexico\\
$^{72}$ Instituto de F\'{i}sica, Universidade Federal do Rio Grande do Sul (UFRGS), Porto Alegre, Brazil\\
$^{73}$ Instituto de F\'{\i}sica, Universidad Nacional Aut\'{o}noma de M\'{e}xico, Mexico City, Mexico\\
$^{74}$ iThemba LABS, National Research Foundation, Somerset West, South Africa\\
$^{75}$ Jeonbuk National University, Jeonju, Republic of Korea\\
$^{76}$ Johann-Wolfgang-Goethe Universit\"{a}t Frankfurt Institut f\"{u}r Informatik, Fachbereich Informatik und Mathematik, Frankfurt, Germany\\
$^{77}$ Joint Institute for Nuclear Research (JINR), Dubna, Russia\\
$^{78}$ Korea Institute of Science and Technology Information, Daejeon, Republic of Korea\\
$^{79}$ KTO Karatay University, Konya, Turkey\\
$^{80}$ Laboratoire de Physique des 2 Infinis, Ir\`{e}ne Joliot-Curie, Orsay, France\\
$^{81}$ Laboratoire de Physique Subatomique et de Cosmologie, Universit\'{e} Grenoble-Alpes, CNRS-IN2P3, Grenoble, France\\
$^{82}$ Lawrence Berkeley National Laboratory, Berkeley, California, United States\\
$^{83}$ Lund University Department of Physics, Division of Particle Physics, Lund, Sweden\\
$^{84}$ Moscow Institute for Physics and Technology, Moscow, Russia\\
$^{85}$ Nagasaki Institute of Applied Science, Nagasaki, Japan\\
$^{86}$ Nara Women{'}s University (NWU), Nara, Japan\\
$^{87}$ National and Kapodistrian University of Athens, School of Science, Department of Physics , Athens, Greece\\
$^{88}$ National Centre for Nuclear Research, Warsaw, Poland\\
$^{89}$ National Institute of Science Education and Research, Homi Bhabha National Institute, Jatni, India\\
$^{90}$ National Nuclear Research Center, Baku, Azerbaijan\\
$^{91}$ National Research Centre Kurchatov Institute, Moscow, Russia\\
$^{92}$ Niels Bohr Institute, University of Copenhagen, Copenhagen, Denmark\\
$^{93}$ Nikhef, National institute for subatomic physics, Amsterdam, Netherlands\\
$^{94}$ NRC Kurchatov Institute IHEP, Protvino, Russia\\
$^{95}$ NRC \guillemotleft Kurchatov\guillemotright  Institute - ITEP, Moscow, Russia\\
$^{96}$ NRNU Moscow Engineering Physics Institute, Moscow, Russia\\
$^{97}$ Nuclear Physics Group, STFC Daresbury Laboratory, Daresbury, United Kingdom\\
$^{98}$ Nuclear Physics Institute of the Czech Academy of Sciences, \v{R}e\v{z} u Prahy, Czech Republic\\
$^{99}$ Oak Ridge National Laboratory, Oak Ridge, Tennessee, United States\\
$^{100}$ Ohio State University, Columbus, Ohio, United States\\
$^{101}$ Petersburg Nuclear Physics Institute, Gatchina, Russia\\
$^{102}$ Physics department, Faculty of science, University of Zagreb, Zagreb, Croatia\\
$^{103}$ Physics Department, Panjab University, Chandigarh, India\\
$^{104}$ Physics Department, University of Jammu, Jammu, India\\
$^{105}$ Physics Department, University of Rajasthan, Jaipur, India\\
$^{106}$ Physikalisches Institut, Eberhard-Karls-Universit\"{a}t T\"{u}bingen, T\"{u}bingen, Germany\\
$^{107}$ Physikalisches Institut, Ruprecht-Karls-Universit\"{a}t Heidelberg, Heidelberg, Germany\\
$^{108}$ Physik Department, Technische Universit\"{a}t M\"{u}nchen, Munich, Germany\\
$^{109}$ Politecnico di Bari and Sezione INFN, Bari, Italy\\
$^{110}$ Research Division and ExtreMe Matter Institute EMMI, GSI Helmholtzzentrum f\"ur Schwerionenforschung GmbH, Darmstadt, Germany\\
$^{111}$ Russian Federal Nuclear Center (VNIIEF), Sarov, Russia\\
$^{112}$ Saha Institute of Nuclear Physics, Homi Bhabha National Institute, Kolkata, India\\
$^{113}$ School of Physics and Astronomy, University of Birmingham, Birmingham, United Kingdom\\
$^{114}$ Secci\'{o}n F\'{\i}sica, Departamento de Ciencias, Pontificia Universidad Cat\'{o}lica del Per\'{u}, Lima, Peru\\
$^{115}$ St. Petersburg State University, St. Petersburg, Russia\\
$^{116}$ Stefan Meyer Institut f\"{u}r Subatomare Physik (SMI), Vienna, Austria\\
$^{117}$ SUBATECH, IMT Atlantique, Universit\'{e} de Nantes, CNRS-IN2P3, Nantes, France\\
$^{118}$ Suranaree University of Technology, Nakhon Ratchasima, Thailand\\
$^{119}$ Technical University of Ko\v{s}ice, Ko\v{s}ice, Slovakia\\
$^{120}$ The Henryk Niewodniczanski Institute of Nuclear Physics, Polish Academy of Sciences, Cracow, Poland\\
$^{121}$ The University of Texas at Austin, Austin, Texas, United States\\
$^{122}$ Universidad Aut\'{o}noma de Sinaloa, Culiac\'{a}n, Mexico\\
$^{123}$ Universidade de S\~{a}o Paulo (USP), S\~{a}o Paulo, Brazil\\
$^{124}$ Universidade Estadual de Campinas (UNICAMP), Campinas, Brazil\\
$^{125}$ Universidade Federal do ABC, Santo Andre, Brazil\\
$^{126}$ University of Cape Town, Cape Town, South Africa\\
$^{127}$ University of Houston, Houston, Texas, United States\\
$^{128}$ University of Jyv\"{a}skyl\"{a}, Jyv\"{a}skyl\"{a}, Finland\\
$^{129}$ University of Kansas, Lawrence, Kansas, United States\\
$^{130}$ University of Liverpool, Liverpool, United Kingdom\\
$^{131}$ University of Science and Technology of China, Hefei, China\\
$^{132}$ University of South-Eastern Norway, Tonsberg, Norway\\
$^{133}$ University of Tennessee, Knoxville, Tennessee, United States\\
$^{134}$ University of the Witwatersrand, Johannesburg, South Africa\\
$^{135}$ University of Tokyo, Tokyo, Japan\\
$^{136}$ University of Tsukuba, Tsukuba, Japan\\
$^{137}$ Universit\'{e} Clermont Auvergne, CNRS/IN2P3, LPC, Clermont-Ferrand, France\\
$^{138}$ Universit\'{e} de Lyon, CNRS/IN2P3, Institut de Physique des 2 Infinis de Lyon , Lyon, France\\
$^{139}$ Universit\'{e} de Strasbourg, CNRS, IPHC UMR 7178, F-67000 Strasbourg, France, Strasbourg, France\\
$^{140}$ Universit\'{e} Paris-Saclay Centre d'Etudes de Saclay (CEA), IRFU, D\'{e}partment de Physique Nucl\'{e}aire (DPhN), Saclay, France\\
$^{141}$ Universit\`{a} degli Studi di Foggia, Foggia, Italy\\
$^{142}$ Universit\`{a} di Brescia, Brescia, Italy\\
$^{143}$ Variable Energy Cyclotron Centre, Homi Bhabha National Institute, Kolkata, India\\
$^{144}$ Warsaw University of Technology, Warsaw, Poland\\
$^{145}$ Wayne State University, Detroit, Michigan, United States\\
$^{146}$ Westf\"{a}lische Wilhelms-Universit\"{a}t M\"{u}nster, Institut f\"{u}r Kernphysik, M\"{u}nster, Germany\\
$^{147}$ Wigner Research Centre for Physics, Budapest, Hungary\\
$^{148}$ Yale University, New Haven, Connecticut, United States\\
$^{149}$ Yonsei University, Seoul, Republic of Korea\\

\bigskip 

\end{flushleft} 
\endgroup  
\end{document}